# U and Th content in the Central Apennines continental crust: a contribution to the determination of the geo-neutrinos flux at LNGS

Coltorti M.[1*], Boraso R.[1], Mantovani F.[2,5], Morsilli M.[1], Fiorentini G.[2,5], Riva A.[1], Rusciadelli G.[3], Tassinari R.[1], Tomei C.[4], Di Carlo G.[4], Chubakov V.[2,5]

1 – Dipartimento di Scienze della Terra, Università di Ferrara
2 – Dipartimento di Fisica, Università di Ferrara
3 – Dipartimento di Scienze della Terra, Università di Chieti
4 – INFN, Laboratorio Nazionale Gran Sasso
5 – INFN, Sezione di Ferrara

[*]Corresponding author:
Via Saragat 1,
Tel.: +39-0532-974721
Fax.: +39-0532-974767
clt@unife.it,

# ABSTRACT

The regional contribution to the geo-neutrino signal at Gran Sasso National Laboratory (LNGS) was determined based on a detailed geological, geochemical and geophysical study of the region. U and Th abundances of more than 50 samples representative of the main lithotypes belonging to the Mesozoic and Cenozoic sedimentary cover were analyzed. Sedimentary rocks were grouped into four main "Reservoirs" based on similar paleogeographic conditions and mineralogy. The initial assumption that similar chemico-physical depositional conditions would lead to comparable U and Th contents, was then confirmed by chemical analyses. Basement rocks do not outcrop in the area. Thus U and Th in the Upper and Lower Crust of Valsugana and Ivrea-Verbano areas were analyzed. Irrespective of magmatic or metamorphic origin lithotypes were subdivided into a mafic and an acid reservoir, with comparable U and Th abundances.

Based on geological and geophysical properties, relative abundances of the various reservoirs were calculated and used to obtain the weighted U and Th abundances for each of the three geological layers (Sedimentary Cover, Upper and Lower Crust). Using the available seismic profile as well as the stratigraphic records from a number of exploration wells, a 3D modelling was developed over an area of 2°x2° down to the Moho depth, for a total volume of about $1.2 \times 10^6$ km$^3$. This model allowed us to determine the volume of the various geological layers and eventually integrate the Th and U contents of the whole crust beneath LNGS.

On this base the local contribution to the geo-neutrino flux (S) was calculated and added to the contribution given by the rest of the world, yielding a Refined Reference Model prediction for the geo-neutrino signal in the Borexino detector at LNGS: $S(U) = (28.7 \pm 3.9)$ TNU and $S(Th) = (7.5 \pm 1.0)$ TNU. An excess over the total flux of about 4 TNU was previously obtained by Mantovani et al. (2004) who calculated, based on general worldwide assumptions, a signal of 40.5 TNU. The considerable thickness of the sedimentary rocks, almost predominantly represented by U- and Th-poor carbonatic rocks in the area near LNGS, is responsible for this difference. Thus the need for detailed integrated geological study is underlined by this work, if the usefulness of the geo-neutrino flux for characterizing the global U and Th distribution within the Earth's Crust, Mantle and Core is to be realized.

# 1. INTRODUCTION

Geo-neutrinos — the antineutrinos from the decay of U, Th and $^{40}$K in the Earth — can provide information on the heat-producing element concentrations of the whole planet. Their detection can shed light on the interpretation of the terrestrial heat flow data, on the present composition and on the origin of the Earth (Rudnick and Gao, 2003; McDonough, 2005).

Geo-neutrino properties, reviewed in (Fiorentini et al., 2007) and summarized in Table 1, deserve a few comments:
(i) geo-neutrinos originating from different elements can be distinguished due to their different energy spectra; e.g. geo-neutrinos with energy E > 2.25 MeV are produced only from the uranium chain;
(ii) geo-neutrinos from U and Th (not those from $^{40}$K) are above threshold for the classical antineutrino detection, the inverse beta reaction on free protons[1]:

$$\bar{\nu} + p \rightarrow e^+ + n - 1.8 \quad MeV \qquad (1.1)$$

(iii) Antineutrinos from the Earth are not obscured by solar neutrinos, which cannot yield reaction (1.1).

Geo-neutrinos were first discussed by Eder (1966) and Marx (1969) soon realized their relevance for geophysics. Raghavan et al. (1998) and Rotschild et al. (1998) pointed out the potential of Kamland (Kamioka Liquid Scintillator Antineutrino Detector), a detector in the Kamioka mine in Japan, and of Borexino, a detector located at the Gran Sasso National Laboratory (LNGS) in Italy, for geo-neutrino detection (Table 2). KamLAND (Araki et al., 2005) presented the first experimental evidence of geo-neutrino production in 2005, and Borexino is at present acquiring data (Bellini, G. et al., 2010). Other experiments aiming at geo-neutrino detection are in preparation (SNO+ at the Sudbury mine in Canada) or in the planning stages (LENA at Pyhasalmi in Finland, Hanohano at Hawaii, Table 2).

A Reference Model (RM) for geo-neutrino production, based on a compositional map of the Earth's crust and on geochemical modeling of the mantle, was presented in Mantovani et al. (2004), with the aim of providing worldwide predictions of geo-neutrino signal.

In principle geo-neutrino measurements can provide quantitative information about the total amounts of U and Th in the Earth and their distribution within the different reservoirs (crust, mantle and possibly core). However, the geo-neutrino signal has a large local component, which depends on the total mass of U and Th in the Earth and on the abundances and distributions of these elements in the region around the detector. For KamLAND and Borexino, Mantovani et al. (2004) estimated that about one half of the signal originates from a volume surrounding the detector with a radius on the surface of 400 km and 800 km, respectively, down to the Moho depth. This region, although containing a globally negligible amount of U and Th, produces a large contribution to the signal as a consequence of its proximity to the detector.

When building the reference model, Mantovani et al. (2004) divided the Earth's crust into 2° x 2° (latitude vs longitude) horizontally homogeneous tiles, following Bassin et al. (2000) and Laske et al. (2001). For each individual tile the model considers the thickness and the density of seven layers: ice, water, soft sediments, hard sediments, upper crust, middle crust and lower crust. Thus the "third dimension" depends on the tile and in general it varies tile by tile. The database can be downloaded from this link: http://igpppublic.ucsd.edu/~gabi/ftp/crust2/. Worldwide averages for the chemical composition of the different regions of the Earth (e.g., upper crust, lower crust, mantle) were used to estimate U and Th concentrations. This is clearly a very rough approximation for describing the region surrounding the detector.

---

[1] This is the reason for considering in this paper only geo-neutrinos from U and Th, not those from $^{40}$K.

If one wants to extract from the total signal relevant information on the deep Earth, the regional contribution to the geo-neutrino flux needs to be determined on the grounds of a more detailed geological, geochemical and geophysical study of the region. The construction of a refined reference model (RRM) for Gran Sasso is the aim of this paper.

In section 5 we present a three dimensional geological model of the 2° x 2° area centered at Gran Sasso National Laboratories (Fig. 1 and 2), down to the Moho depth, based on the results of a deep seismic exploration of the Mediterranean and Italy (the CROP project) (Finetti, 2005a), as well as geological and stratigraphical distribution of the sedimentary covers (SC) recognizable from geological maps, integrated with data from deep oil and gas wells. For this tile a detailed 3D model was performed where the thickness of the sediments, upper and lower crust layers change point by point. The main feature of this area is a thick sedimentary cover, which was not adequately accounted for in the averages leading to the 2° x 2° crustal map of Mantovani et al. (2004).

When building the reference model, Mantovani et al. (2004) used average abundances that were based on measurements of worldwide collections of samples. To check that these global averages are appropriate for the Gran Sasso region, we analyzed representative samples of the sediments and upper and lower crustal lithologies in Northern Italy (Ivrea-Verbano zone and Valsugana). The results of this study, presented in section 6, confirm the adequacy of the world-wide averages adopted for the reference model.

In section 7 we compute the regional contributions to the geo-neutrino signal according to the more refined model and compare it with previous estimates. The concluding section summarizes our results.

## 2. THE REFERENCE MODEL FOR GEO-NEUTRINO SIGNAL AT GRAN SASSO

A reference model for geo-neutrino production is a necessary starting point for studying the potential and expectations of detectors at different locations. By definition, it should incorporate the best available geological, geochemical and geophysical information on our planet. In practice, it has to be based on selected geophysical and geochemical data and models (when available), on plausible hypotheses (when possible), and admittedly on arbitrary assumptions (when unavoidable) (Fogli et al., 2006). Recently a few such models have been presented in the literature (Mantovani et al., 2004; Fogli et al., 2006; Enomoto et al., 2007). Predictions by different authors for a few locations are compared in Table 2, where the expected geo-neutrino signal is expressed in Terrestrial Neutrino Units (1 TNU corresponds to one event per $10^{32}$ target protons occuring at the detector per year of exposure time).

All these models rely on the geophysical 2° × 2° crustal map of Bassin et al. (2000) and Laske et al. (2001) and on the density profile of the mantle as given by PREM (Dziewonski and Anderson, 1981). Adopted U and Th abundances for the various layers in Mantovani et al. (2004) are shown in Table 3. The abundances in the crustal layers were obtained by averaging results that were available on the GERM (http://earthref.org/) database in 2002. A chemically layered mantle was assumed, with U and Th abundances in the upper mantle from Jochum et al. (1983), Zartman and Haines (1988), Salter and Stracke (2004) and Workman and Hart (2005). The Bulk Silicate Earth mass constraint was used in order to determine the abundances in the lower portion of the mantle.

The last column of Table 4 shows the contributions of the different reservoirs to the geo-neutrino signal at Gran Sasso, according to the reference model of Mantovani et al. (2004). The mantle contributes 9 TNU, about 20% of the total signal, while crust and sediments all over the world provide the rest (31.5 TNU). Half of this originates from the six tiles depicted in Fig. 1 which provide a "regional crustal contribution":

$$S_{reg} = 15.3 \quad TNU \qquad (2.1)$$

Within this region, the 2° x 2° Central Tile, indicated as CT in Fig. 1, generates a "local contribution" of:

$$S_{CT} = 11.8 \quad TNU \qquad (2.2)$$

All this demonstrates the particular importance of the region close to the detector, which warrants a closer look. Geological units and structures present in the detector area, which might be washed out in the 2°×2° crustal map, have to be considered. In addition, the differences in the geochemical composition of different reservoirs surrounding the detector compared to the world averages need to be evaluated.

The regional flux needs to be determined with an accuracy that is comparable to uncertainties from the contributions of the rest of the Earth. This is, in its essence, the rationale for building a refined reference model (RRM).

Before closing this section, it is useful to take into account the position of the LNGS, which in Mantovani et al. (2004) was rounded to 42° N, 13° E. This approximation was adequate given the tile size of the 2°x2° crustal map used in that reference model. For the refined reference model, a better precision is required. The geographical position of the underground laboratory is 42° 27' N and 13° 34' E of Greenwich, see Bellotti (1988); this more precise position is adopted here. By changing the position to this value, holding all other parameters constant, the predicted signal increases by 0.5 TNU, i.e. about 1%, (Table 4).

### 3. SAMPLING AND ANALYTICAL METHODS

In order to check if the values for the geo-neutrino flux adopted by Mantovani et al. (2004) in the RM are appropriate for the Gran Sasso area, we carried out an accurate sampling of the sedimentary rocks within a distance of 20km of LNGS. Chemical homogeneity of the formations were further constrained on the basis of a less extensive sampling over a larger area from Gargano Promontory to Ancona (Fig. 1). Upper and Lower Crust (UC and LC respectively) material does not outcrop around the studied area. Samples were thus taken from the LC-UC section of Ivrea-Verbano Zone from granulitic rocks of Val Strona, Val Sessera and Val Sesia to amphibolitic schists of Serie dei Laghi and related intrusives (Hunziker and Zingg, 1980; Borghi, 1988; Boriani et al., 1990; De Marchi et al., 1998; Quick et al., 1992; 2003; Franz and Romer, 2007). To complete the UC section, the low grade metamorphic philladic rocks outcropping in Valsugana, and related intrusive rocks of Caoria and Cima d'Asta complexes were also sampled (Dal Piaz and Martin, 1998; D'Amico et al., 1971; Sassi et al., 2004). This ex situ sampling was carried out assuming that rock abundances and the composition of the south Alpine basement are fairly homogeneous for the whole Adriatic microplate.

Taking into account the considerable thickness (>10 km) of the sedimentary cover around Gran Sasso area we prefer to treat it and the Upper Crust separately. In contrast, due to the geological and geophysical difficulties of defining precisely the intermediate layer introduced by Rudnick and Fountain (1995) we prefer to subdivide the crust in only two layers, namely the Lower Crust and the Upper Crust, as also proposed by Wedephol (1995).

In summary for the sedimentary successions 28 samples, 14 within 20 km and 14 within 200 km from the LNGS were collected (Table 5). They are considered representative of the principal geological units outcropping in the region, according to Vezzani and Ghisetti (1998) geological map. 29 metamorphic and intrusive samples were collected in Valsugana and Ivrea-Verbano-Laghi areas, with particular emphasis to felsic rocks of the UC due to their high content in radionuclides (Table 6A and B).

All samples were fresh, without any visible chemical alteration. Representativity was enhanced by taking large samples (>2 kg) which were sliced and ground almost entirely. Th and U abundances were determined by Inductively Coupled Plasma-Mass Spectrometry (ICP-MS) using a VG Elemental Plasma Quad PQ2 Plus, at the Department of Earth Sciences, Ferrara University. Accuracy and precision were calculated by analyzing a set of international rock standards with certified values from Govindaraju (1994). These geostandards include: JP-1, JGb-1, BHVO-1, UBN, BE-N, BR, GSR-3, AN-G, MAG-1, JLs1 and JDo1. Accuracy for analyzed elements is in the range of 0.9–7.9 relative %. Conservatively, we shall assume an accuracy of 10%. Detection limits for U and Th are 0.01 ppm.

The analytical procedure begins with the dissolution of about 200 mg of rock powder into 50 mL PTFE beaker with 3 ml HNO3 65% (Suprapur® Merck) and 6 ml HF 40% (Suprapur® Merck). The beaker is covered by Parafilm and put in an ultrasonic bath for about 15 minutes. After at least 12 hours the Parafilm is removed and the sample evaporated to incipient dryness on hot plate at about 180° C. 3 ml of HNO3 65% and 3 ml of HF 40% are subsequently added to the beaker and the sample is further evaporated to incipient dryness. The complete removal HF is realized by evaporation with 4 ml $HNO_3$. Finally the sample is taken into 3 ml $HNO_3$ and transferred into 100 ml polypropylene volumetric flasks. Solutions of Rh, In, Re, Bi are eventually added as internal standards to the flask that are then made-up to volume.

Some samples have also been measured through gamma spectrometry by means of a 3'x3' NaI(Tl) crystal from ORTEC, installed in an underground building inside Hall A of the LNGS (Arpesella et al., 1996). This unique location guarantees a reduction factor of the cosmic ray flux of about one million (Bellotti, 1988). The detector was enclosed in lead shielding 15 cm or more thick on all sides, in order to shield against the residual natural radioactivity of the environment. The powder samples were kept in cylindrical plastic boxes of approximately 50 $cm^3$ volume, the rock samples were kept in similar boxes of larger volume, according to their sizes. To evaluate the counting efficiency of our detector we have used a Monte Carlo simulation program based on the Geant4 code (Agostinelli et al., 2003), widely used in the fields of high-energy, astroparticle and underground physics. Detection limits for U and Th are at the level of 0.1 and 0.3 ppm respectively. Measurement errors are estimated by taking into account: i) number of counts in the gamma peak, ii) number of counts in the background peak and iii) counting efficiency. Measurements errors are thus estimated for each sample. For a more detailed description of this technique see Appendix 1.

In summary, ICP-MS has a detection limit which is an order of magnitude lower than that of the NaI counting system; also, the instrumental error comes out to be smaller for elemental abundances up to few ppm. On the other hand, the NaI method, being more direct with respect to the treatment of the samples, does not suffer of uncertainties about the effectiveness of the chemical attack. It has to be noted, however, that in the NaI method one measures the intensity of lines from the daughters in the decay chain and the abundances of the parent element can only be inferred under the assumption of secular equilibrium, an hypothesis which is avoided with the ICP-MS method.

For most of the samples, U and Th mass abundances have been measured with both methods, generally with consistent results (Tables 5 and 6A and B).

A further control on Th abundance was also performed by using X-Ray spectrometer on pressed powder pellets, using an ARL Advant-XP spectrometer and following the full matrix correction method proposed by Lachance and Traill (1966). The detection limit is estimated to be 1 ppm. Analytical accuracy is better than 10 %. It is reported as the average of the relative differences between measured and recommended values in 18 reference standards (including igneous, metamorphic and sedimentary rocks) with Th concentration ranging from 2.6 to 106 ppm. Analytical precision, as relative standard deviations of replicate analyses (six in two year's elapsed time), is usually < 15%. For the sake of clarity these data are not reported in the tables, but they agree well with the other two sets.

For sediment samples, we adopted the values provided by the more precise ICP-MS, using NaI method as a check. Results are statistically consistent except for the U measurement of sample 08MM, which was rejected (Table 5).

For crust samples the two methods provide consistent results, both for felsic and mafic rocks. For those samples whose Th and U concentrations are below or near the detection limit for NaI, the ICP-MS values were adopted. When the accuracy of the two methods are comparable the error weighted average of the two results are adopted. U concentrations measured with NaI in felsic rocks of the UC are systematically higher than those derived with ICP-MS, except for sample VS9 (Table 6A). This might be related to incomplete dissolution of zircons, which are abundant in these rocks, notwithstanding the clear aspect of the solution. For these rocks the NaI values were adopted.

## 4. THE GEOLOGY FRAMEWORK OF CENTRAL ITALY

Apennines, together with other peri-mediterranean mountain chains (Southern Alps, Dinarids and Ellenids) and their related continental crust, were part of the Adria plate (see Cavazza and Wezel, 2003). This plate is seen as an independent microplate (Anderson and Jackson, 1987; Oldow et al., 2002; Battaglia et al., 2004) or as a promontory of the African shield (Channel et al., 1979; Babbucci et al., 2004) whose evolution resulted from a complex geological history, punctuated by several events related to the evolution from divergent to collisional continental margins occurred in the Mediterranean region, starting from the early Mesozoic (Dewey et al., 1973; Jolivet and Facenna, 2000; Wortel and Spakman, 2000; Carminati et al., 2005; Lucente et al., 2006; Panza et al., 2007; Mantovani et al., 2009; Viti et al., 2009). With time geodynamical processes have produced the geological structure of the Central Apennines where the LNGS is located (Fig. 2) (Ghisetti and Vezzani, 1999; Bigi and Costa Pisani, 2005; Finetti et al., 2005a; Scisciani and Calamita, 2009).

The Apennines developed through the deformation of two major paleogeographic domains: the Liguria-Piedmont Ocean and the Adria-Apulia passive margin (Bernoulli, 2001; Elter et al., 2003; Bosellini, 2004; Parotto and Praturlon, 2004) which were progressively subducted below the European plate or incorporated into the chain during the geodynamic events lasting from Late Cretaceous to Plio-Pleistocene (Scisciani and Calamita, 2009 and references therein). Actually, the northern and central Apennines are an arc shaped fold-and-thrust belt, with north-eastward convexity and vergence that plunges north-westward (Barchi et al., 2001).

We have modelled an area of 2° x 2° of latitude and longitude, centred on the LNGS (Fig. 2). This area includes the following main geological domains of central Italy:
1) The northern Apennines, bounded southward by the Olevano-Antrodoco Line;
2) The central Apennines, bounded southward by the Sangro-Volturno Line;
3) The external Apennine foredeep developed at the front of the two main structural arcs;
4) The peri-Adriatic foreland, developing externally to the main thrusts fronts and in the Adriatic offshore.

Gran Sasso Range (GSR), where the LNGS is located, represents the northernmost front of the Abruzzi Apennines (Scisciani et al., 2002; Sani et al., 2004; Scrocca et al., 2005; Billi and Tiberti, 2009). Structural elements are represented by a complex system of overturned anticlines and related thrusts (Vezzani and Ghisetti, 1998; Ghisetti and Vezzani, 1999; Speranza et al., 2003; Calamita et al., 2006). The result of these movements is a juxtaposition of the northern margin of the Lazio-Abruzzi carbonate platform and its related pelagic Umbria-Marche Basin onto the external Apenninic foredeep (Bernoulli, 2001; Bosellini, 2004; Parotto and Praturlon, 2004; Finetti et al., 2005a).

Based on the geodynamic and structural evolution of this area, the sedimentary pile of the GSR has been subdivided into three main sequences. The first two sequences are related to the syn- and postrift evolution of the Adria paleomargin, and range from Late Triassic to Late Paleogene (Bernoulli, 2001; Bosellini, 2004; Parotto and Praturlon, 2004; Finetti et al., 2005a). The third sequence relates to the progressive deformation of the Adria paleomargin, connected with the building process of the Apenninic chain (foreland to thrust-top basins) during the Late Paleogene to Pleistocene (Mostardini and Merlini, 1986; Argnani and Frugoni, 1997; Cipollari et al., 1999; Patacca and Scandone, 2001, 2007).

The lithostratigraphic framework is defined by different types of sedimentary successions reflecting various depositional environments (Parotto and Praturlon, 1975; 2004). Carbonate systems develop during the Mesozoic and early Tertiary with two main depositional systems: carbonate platforms characterize the Early Triassic to Late Cretaceous syn- to post-rift evolution, whereas carbonate ramps dominate the transition from the post-rift to the early stages of the foreland evolution (Eberli et al., 1993; Bernoulli, 2001; Bosellini, 2004; Parotto and Praturlon, 2004). Silicoclastic depositional systems, mainly characterize foredeep evolution of the central Apennines, with terrigenous deposits progressively overlying the previous carbonate depositional systems, from west to east, starting from the late Miocene onward (Cipollari et al., 1999).

**4.1 – Sedimentary Cover**
A Geological map of the area (Fig. 2) clearly shows a broad division in four main lithofacies: shallow-water carbonate, relatively deep-water carbonate, siliciclastic deposits related to the foredeep, chaotic complexes. They are composed by numerous sedimentary units whose thickness, age and depositional environment are reported in Table 7, together with reservoir classification (introduced at Chapter 6) and label of the samples taken for this study (Table 5).

In the lower left-hand corner of the map the volcanic products of the Roman Magmatic Province also outcrop (Conticelli et al., 2009). Taking into account the distance from the detector and the volumetric abundance of these rocks with respect to the huge pile of sediments within the Sedimentary Cover the influence of these rocks on the geo-neutrino flux has been neglected.

*Permian – Early Triassic* - the sedimentary succession starts with a continental clastic deposits (Verrucano Auct.) (Bernoulli, 2001; Vai, 2001) mainly composed by conglomerate and sandstones (Reservoir b3). In some area a Permian shallow-marine carbonate succession also occurs (see Gargano 1 and Puglia 1 wells in Apulia).

*Late Triassic - Early Jurassic* – During this time an evaporitic basin develops, with the deposition of the "Anidriti di Burano" (Martinis and Pieri, 1964), a widespread formation found in the northern Apennines and in the southern Italy (Zappaterra, 1994; Ciarapica and Passeri, 2002) (Reservoir b2). In the Abruzzi and Latium regions coeval deposits are represented by shallow-water carbonate known as "Dolomia Principale" (Norian to Rhaetian).

*Early Lias* – Shallow water carbonate sedimentation persists in all the sectors with the deposition of the "Calcare Massiccio" Fm. until the Hettangian-Sinemurian when the rifting phase of the future Alpine Tethys starts (Parotto and Praturlon, 1975) (Reservoir b2). This phase starts with the timetransgressive drowning of the Umbria-Marche paleogeographical domain (Montanari et al., 1989; Santantonio, 1993) and the developments of areas with condensed (seamounts) and normal pelagic sequences (Coltorti and Bosellini, 1980) ("Rosso Ammonitico", "Corniola", "Bugarone", "Maiolica", "Scisti a Fucoidi", "Scaglia Bianca", "Scaglia Rossa", "Scaglia Cinerea") (Reservoir b1).

*Middle - Late Lias* – In some areas shallow-water carbonate deposits persists with the deposition of Calcari a Palaeodasycladus Fm, while in some others slope sediments developed (megabreccias and calciturbidites).

*Dogger - Malm* – In the areas occupied by carbonate platforms oolitic and micritic limestone are abundant ("Morrone di Pacentro" Fm, "Monte Acquaviva" Fm). In the basinal areas there are the typical facies with widespread formation as "Calcari a Filaments", "Diaspri", "Marne ad Aptici" and the lower part of the "Maiolica" Fm.

*Early Cretaceous* – Carbonate sedimentation continuous with small hiatus until late Albian to middle Cenomanian. In the whole basinal areas the deposition of typical pelagic facies take place with the "Maiolica" and "Marne a Fucoidi" Fms: in this latter formation some black-shales layers occur related to the oceanic anoxic events (Jenkyns, 2003) (Reservoir b1 and b2).

*Late Cretaceous* – In this period there is the development of the carbonate platforms dominated by large bivalve ("Calcari a Rudiste", "Orfento" Fms.) whereas pelagic lithotypes accumulate in basinal area with the "Scaglia" Fm ("Scaglia Bianca" and "Scaglia Rossa") (Eberli et al., 1993) (Reservoir b2).

*Paleogene* – The sedimentation become discontinuous with various lacuna in different platform areas. In the basinal area there are the pelagic successions with the Scaglia Cinerea Fm (Reservoir b1).

*Miocene* – After the widespread lacuna of Paleogene a new marine transgression develop a carbonate ramp ("Calcari a Briozoi e Litotamni" Fm) successively covered by terrigenous deposits of the Apennine foredeep (Laga Fm., Reservoir a; Cipollari et al., 1999).

*Pliocene and Pleistocene* - In the Adriatic sector the middle-Pliocene transgression cover unconformably the Messinian or lower Pliocene sediments, with basal conglomerate, clays and sands (Laga Fm., Reservoir a). The marine Pleistocene follows these deposits with clayey-sandy sediments with thickness reaching up to 1000 m ("Cellino" Fm; Carruba et al., 2006).

## 4.2 – Upper Crust

The Upper Crust (UC) can be defined as the portion of continental crust ranging from the bottom of the sedimentary cover to the Conrad discontinuity, which marks the top of the Lower Crust. This approach, which is mainly based on geophysical data, shows some problems when trying to define the position of the medium-grade, amphibolite facies rocks. Indeed, Rudnick and Fountain (1995) introduced a Middle Crust mainly composed by migmatitic rocks, whereas Wedepohl (1995) includes the amphibolites within the UC. The geophysical profiles beneath Central Italy do not support the existence of a well defined intermediate level, while the limited thickness of the LC induce us to insert the amphibolitic rocks within the UC.

Two classical outcrops of the Southern Alps were sampled as representative of the UC: the Serie dei Laghi near Lago Maggiore, and the phillacic basement of Valsugana, near Trento (Table 6A). The Serie dei Laghi outcrop in the SE portion of the Ivrea-Verbano zone. It corresponds to intermediate to upper crust lithologies. The deepest part is made up of Ordovician meta-sandstones to metapelites (paragneiss with calc-silicate inclusions and biotite- and muscovite-bearing gneiss) in amphibolitic facies intruded by dioritico-granitic orthogneiss with calc-alkaline affinity (Borghi et al., 1991). In the upper part micaschists and paragneiss predominates with two micas, garnet, staurolite and kyanite sometime retrograded to green-schist facies.

In the Valsugana area the pre-Hercynian sediments are slightly metamorphosed. It consists of two phillitic units separated by a complex volcano-sedimentary series with meta-carbonates and metarhyolites. These metamorphic terrains are intruded by calc-alkaline plutons varying in composition from diorite to granodiorites and granites (prevalent) (Cima d'Asta and Caoria plutons; D'Amico et al., 1971).

**4.3 – Lower Crust**

LC rocks are available only thanks to tectonic processes which denudate the deepest portion of the crust or to basaltic magmatism scavenging small pieces of LC (xenoliths) during their uprising from the mantle to the Earth's surface. The two geological evidences however lead to contrasting results as far as relative proportions of mafic and felsic rocks are concerned. As reviewed by Rudnick and Fountain (1995), mafic rocks are more represented in xenoliths, than usually are in tectonically emplaced LC terrains where, on the contrary, felsic rocks are equally encountered if not predominant. Horizontal heterogeneity within LC is of course expected, but the simplest explanation for this contrast may lie on the tool we are using for sampling. Where basalts are present, underplating process may have been working even for a long while, thus mafic rocks in this settings - where xenoliths are taken - may be more extensively represented. An example of this situation may be seen in the Ivrea Zone which, as already remarked, represents the LC outcrop nearest to the investigated area. In Val Sessera and Val Sesia in fact gabbroic rocks predominate as a result of various intrusions, several km in thickness, while in Val Strona felsic rocks are more abundant.

Due to the lack of LC outcrops in the area close to Gran Sasso, representative LC samples were taken from the Ivrea-Verbano Zone which represents the most classic and extensive deep crustal section of the Alps. It can be distinguished in two main lithological units: a) the layered complex Permian in age (278-280Ma) in contact with the Canavese Line with thickness up to 10 km in Val Sesia and Val Sessera (Quick et al., 1992; 2003 and reference therein) and b) the Kinzigitic Fm. The former unit is made up of layered gabbros intruded in the deep crust and re-equilibrated in granulite facies (Table 6B). Magmatic lithotypes varies from cumulitic peridotites, pyroxenites, gabbros, anorthosites and diorites (Rivalenti et al., 1980). The intrusion of this body occurs within the older Kinzigitic Complex in a general extensive regime which allow the creation of progressive enlarging magma chamber/s. The heat flux generating by the incoming basic magmas most probably caused the partial melting of the lower crust lithologies (crustal anatexis) which ultimately generated the acid intrusive and effusive magmas of the Serie dei Laghi (Table 6A). Few slices of subcontinental mantle occur at the base of the gabbroic intrusion near the tectonic contact with the Canavese Line (Balmuccia, Baldissero and Finero ultramafic complexes; Rivalenti et al., 1980; Coltorti and Siena, 1984; Mazzucchelli et al., 1992). The Kinzigitic Fm. is constituted by meta-pelites with biotite, sillimanite and garnet intercalated with meta-basites with tholeiitic affinity, marble, calc-silicatic fels and rarely Mn-bearing quarztites. These rocks show Variscan metamorphism, varying from anphibolitic to granulites facies moving northwestward (Valle Strona), predating the gabbroic intrusives (Table 6B).

# 5. THE 3D GEOLOGICAL MODEL OF THE GRAN SASSO AREA

## 5.1 – The Central Tile

A simplified three dimensional (3D) model has been developed on an area of 2° x 2° of latitude and longitude, centred on the LNGS (Central Tile) (Figs. 1 and 2). The upper boundary of the model was fixed at the mean sea-level whereas the lower boundary was fixed at the Moho discontinuity, based on the Moho isopachs map of Finetti et al. (2005b).

The starting points for model building are the profiles published by the CROP Project (Finetti, 2005a). The CROP (CROsta Profonda, i.e. deep crust) is an Italian project for the study of the crustal structure by means of near-vertical reflection seismics. This method is the most used in hydrocarbon exploration, and it has been adapted to reach crustal depths. The CROP Project is similar to other seismic-based deep crust studies, such as the COCORP (USA), DEKORP (Germany), ECORS (France), and BIRPS (England). Although various models have been recently proposed (Cavinato and De Celles, 1999; Billi et al. 2006, Di Luzio et al. 2009) we preferred to follow Finetti's model because of the more abundant information in the sections. The Moho surface was extracted by digitizing the Moho Isopachs Map (CROP) for both Adriatic and Apenninic crustal blocks, originally obtained from seismic and gravity data (Finetti, 2005b). The Moho surface of the Adriatic micro-plate required a propagation of the data under the Apenninic crust because of the absence of information in the Moho Isopachs map. The propagation required the application of a kriging interpolation corrected with cubical drift because the non stationarity of the surface. This propagation was also used to check the base of the crust with respect to the available crustal sections published in Finetti et al. (2005b).

A simplified tectonic model was applied inserting only the main crustal thrusts (Northern Apenninic thrust and Olevano-Antrodoco Lines, Gran Sasso and Southern Apenninic thrust) that are reported in Finetti et al. (2005a) and Calamita et al. (2006). The Northern Apenninic and the Southern Apenninic thrusts are also known as "Moho doubling Thrust" (Calamita et al., 2006) and are separated by an oblique thrust ramp (Olevano-Antrodoco Line). The surface path of these main faults was traced using the Structural Model of Italy (Bigi et al., 1992). In this tectonic framework we have populated the model with the six layers defined below (see chapter 6). A complete highresolution geological and tectonic model would be too great to handle and surely over dimensioned with respect to the aim of the present work.

The conversion from seismic travel-times to actual depths (Finetti, 2005b) was performed going upward, starting from the constructed Moho surface. This solution was preferred because of the large uncertainties for the velocities in the sedimentary cover as explained by Finetti (2005b). It also provides more constant velocities for the Upper and Lower Crust with respect to the more variable velocities recorded within the Sedimentary Cover. In fact, we used graphical representations of the crustal seismic sections, with attendant conversion uncertainties. In particular, the largest uncertainties were concentrated in the upper part of the crust (Sedimentary Cover), where we used public domain data coming from hydrocarbon explorations. Geophysical profiles were then crosschecked with position and depth of the various formation obtained from 53 exploration wells (Mostardini and Merlini, 1986). The main inputs used for building the 3D model are summarized in Fig. 3.

A 3D grid was then constructed by using Schlumberger-Petrel software, in order to quantify the bulk rock volumes of the six main layers. These model surfaces were also modelled using kriging interpolation with a cubical drift. For the model input we used more than 1000 points for the base of the grid and more than 250 points in the interior of the grid. The faults were modelled using more than 1000 points. The numerical output of the model is a file which contains, for each cell the latitude, longitude and depth of its center, volume of the cell and reservoir type. The grid has 1.1 x $10^6$ cells, each with a volume of about 2 $km^3$. The typical size of each cell is 2 x 2 x 0.5 km. The resulting model is illustrated in four simplified geological sections (Fig. 4), which were built in order to satisfy and to cross-check with the CROP published models and interpretations on the Central Apennines of Finetti (2005a). Some incongruence however between the geological map and these sections may be noted. For instance the thin volcanic cover of the Alban Hills, together with some other subordinate geological features (e.g. Montagna dei Fiori) are not reported. Although significant from the geological point of view, these structures in fact are negligible for the calculation of the geo-neutrino flux.

After building the model, we calculated the total volumes for each crustal unit. From Table 8 it is evident that ca. 80% of the total volume of the sedimentary cover is given by the Permo-Mesozoic succession, the largest fraction being composed by the Mesozoic carbonate units. In contrast, Wedepohl (1995) estimates that about 40% of the mean European sedimentary cover consists of carbonates.

It is interesting to compare the thickness of the different layers in the present model and in the crustal model that was used for the reference model (Table 8). The sediment layer is over 25 times thicker in the refined reference model than assumed in the previous crustal model, whereas the Moho depths are within ten percent.

The main uncertainties in defining the points used for building the model come from velocity-depth conversion, which is critical for the best accuracy of the grid. We tried to estimate the uncertainties using the velocity-depth conversion using all available data in the literature about seismic velocities in the crust. The estimated depths of individual reservoirs are dependent on the model that is being used, whereas the Moho depth between the different models is within 15%.

**5.2 – The Rest of the Region**

For the rest of the region – i.e., what remains of the six tiles after subtracting the central tile (Fig. 1) – we performed a less detailed study, since this area is expected to contribute a much smaller fraction of the signal. We distinguished three layers (sediments, upper and lower crust) and we used the following ingredients:

1) Moho depth is taken from the map of Finetti (2005b).
2) The three CROP sections (n. 3 Pesaro–Pienza, n. 4 Barletta–Acropoli, n. 11 Pescara – Civitavecchia, and M2A) are used to build 29 virtual pits, extending from the surface to the Moho.
3) A structural axis, NW-SE (Bigi et al., 1992), corresponding to the merging between the Adriatic and European plates, was identified.
4) Information on the depth of each layer was obtained by linearly interpolating the values available on adjacent CROP lines along the structural axis.

In this way, the depth of different layers was estimated on a mesh of 1/4° x 1/4°. A representative view is shown in Fig. 5.

## 6. Th AND U RESERVOIRS IN THE CENTRAL TILE

The geochemistry of the various lithotypes making up the SC, UC and LC are reported in Tables 5, 6A and 6B. In the following the chemical composition of each individual reservoir as well as their relative abundances in constituting the three main geological meaningful layers (SC, UC and LC) will be calculated.

**6.1 – Sedimentary cover**

We shall assume that sediments formed in similar depositional environments would have similar and rather homogeneous chemical characteristics, thus linking geochemistry to lithofacies. In this respect our approach is similar to that proposed by Plank and Langmuir (1998), allowing to reduce the number of samples to be analysed but preserving at the same time the geologically meaningful information. They estimate that even a reduced number of samples introduces an error in the geochemical estimates of <30%, which considering the aim of this work and the large variety of sediments is a favourable result. This approach is not exhaustive but, taking into account the overall approximation (see also Appendix 2) seems adequate for estimating the geo-neutrino fluxes.

For the purpose of this work the Cenozoic terrigenous and the terrigenous/carbonatic Permo-Mesozoic succession has been divided in four reservoirs (Table 7):

a) – Cenozoic terrigenous units (sandstones, siltites and clays)
b1) – Meso-Cenozoic basinal carbonate units (marly and shaly carbonates, sometimes with black shales)
b2) – Mesozoic Carbonate Units (limestones, dolomites and evaporites, with a negligible marl and clay content)
b3) - Permian clastic units (sandstones and conglomerates)

U and Th mass abundances in the three reservoirs are obtained by averaging arithmetically data for the samples analyzed (Table 8). Lithotypes belonging to the last reservoir (b3) outcrop rarely within the entire Italian Peninsula and due to their conglomeratic nature, sampling is quite difficult. For these reasons and taking into account that the Permian clastic units ("Verrucano" Fm.) result from the dismantling and erosion of the Paleozoic basement rocks, we assume U and Th contents of reservoir b3 are similar to those of the UC.

In each reservoir, the dispersion of the measured abundances is much larger than the analytical uncertainty and also the uncertainties of the mass of the reservoirs are negligible with respect to them. The uncertainty quoted in Table 5 is the standard deviation among the different samples weighted with the mass of the reservoirs. Taking into account the relative volume (Table 9) of the four reservoirs estimated on the base of the 3D geological model (see below) the abundances of U and Th for the whole sedimentary cover can be determined (Table 8).

The largest area in the Gran Sasso region is occupied by U- and Th-poor Mesozoic carbonates. The mass weighted average concentrations are U = 0.8 ppm and Th = 2.0 ppm, which are significantly lower than the world average for sediments adopted in the reference model, U = 1.7 and Th = 6.9 ppm (Plank and Langmuir, 1998). This is a consequence of the large fraction of carbonates in the sedimentary cover. Indeed, the LNGS is located inside a U- and Th-poor carbonate mountain. Two exploratory drill cores made at the time of excavation for the detector provided an opportunity to measure U and Th abundances of the rocks in the tunnel by means of a GeLi instrument (Campos Venuti et al., 1982). Thirty samples were collected and analysed, with the result $\alpha(U_{eq}) = (0.12 \pm 0.11)$ ppm and $\alpha(Th_{eq}) = (0.45 \pm 0.16)$ ppm (see Appendix 1 for the notation of $U_{eq}$ and $Th_{eq}$).

**6.2 – Upper Crust**

In the last two decades a great effort has been dedicated in order to infer crustal composition as a function of depth by comparing the results of seismic profiles with high-pressure laboratory measurements of seismic velocity for a wide range of rocks (Holbrook et al., 1992; Christensen and Mooney, 1995; Rudnick and Fountain, 1995; Wedepohl, 1995; Gao et al., 1998; Behn and Kelemen, 2003).

We note that, although the compressional wave velocity depends on many factors (temperature, mineralogical composition, confining pressure, anisotropy and pore fluid pressure), felsic rocks are characterized by sound speed generally lower than in mafic rocks. Two groups of samples were thus defined: one felsic and another intermediate/mafic (Tables 6A). Average elemental abundances for the two groups were calculated and seismic arguments used in order to fix their relative amounts within UC.

The collected felsic rock types (granite, granodiorite, quartz schist and felsic gneiss) are characterized by compressional wave velocity $v_p(f)$ near 6.2 km/s, while intermediate/mafic rock types (amphibolite, diorite and gabbro) have $v_p(m)$ close to 6.8 km/s. These values for $v_p$ refer to a depth of about 15 km, assuming a stable geotherm (15°C/km) (Holbrook et al., 1992; Christensen and Mooney, 1995).

Several authors have investigated the deep structure of the central Apennines, in particular analysing data from the eastern part of CROP 11 (Finetti, 2005b; Cassinis et al., 2005; Patacca et al., 2008; Di Luzio et al., 2009). For this area the comparison among different estimates, yields for the upper crust:

$$v_p = 6.32 \pm 0.30 \quad km/s$$

From these data we can deduce that the upper portion of the crust of the central Italy is prevalently felsic. The fraction f of felsic rocks and that of mafic rocks ($m = 1 - f$) can be determined by requiring that the observed value of vp in the crust is reproduced, i.e. $f = [v_p(m)-v_p] / [v_p(m)-v_p(f)]$. This gives:

$$f = 0.75 \pm 0.40$$

Values >1 are meaningless from a geological point of view. This is due to the large uncertainties in $v_p$, whose lowest value (6.02 km/s) results lower than the experimentally determined $v_p(f)$. Nevertheless, this result is consistent with the composition of crustal models available in the literature. Christensen and Mooney (1995) assigned 75% of felsic and 25% of mafic rocks to the crustal depth between 10 and 25 kms. Wedephol (1995) identifies nearly 85% of felsic and 15% of mafic reservoirs for the (sediment-free) UC layer and marks the transition UC/LC with the 6.5 km/sec discontinuity in the European Geotraverse. Rudnick and Gao (2003) distinguish an upper from a middle crust. Their middle crust include rocks on amphibolitic facies. They describe several exposed middle crust cross-sections worldwide, where the felsic reservoir generally predominates (90%).

At this point the elemental abundances a for the whole Upper Crust can be calculated, from $a = a_f f + a_m (1-f)$, where $a_f$ and $a_m$ represent the U or Th abundances in the felsic and mafic reservoirs respectively, which in turn result from the arithmetical average. Again, the dispersion among the samples is larger than the measurement errors. The results are shown in Table 10, where comparable uncertainties arise from the spread of elemental abundances among rock types and from the uncertainty on the felsic/mafic percentage.

The range of published values for U and Th abundances in the UC (last line of Table 10) essentially overlaps with the range provided by our estimates. These latter have no pretension of being more accurate: by using material collected in a region relatively close to central Italy, we have provided a check of other studies, based on worldwide samples.

**6.3 – Lower Crust**

Analyses reported in Table 6B show that U abundance is quite low in all granulites (0.01-1.14 ppm), while the average Th content in felsic granulites is more than one order of magnitude higher than in the mafic (see also Schnetger, 1994). Values for the four samples of mafic rocks and for the five samples of felsic lower crustal rocks have been averaged and results are reported in Table 11.

Once again, however the relative proportions of the two components needs to be determined by using indirect methods.

Following seismic arguments, sound speeds in the range 6.7-7.2 km/s have been detected in the area of interest (Ponziani et al., 1995; Finetti, 2005b; Cassinis et al., 2005; Mele et al., 2006; Patacca et al., 2008; Di Luzio et al., 2009). This range appear to be in agreement with the results Christensen and Mooney (1995) and Holbrook et al. (1992) who measured an average p-wave velocity of 6.3 and 7.0 km/s for the felsic and mafic reservoirs respectively. If the lowest value is considered a felsic/mafic proportion of ca. 40/60 can be obtained, whereas the highest value would bring the composition of the LC completely composed by mafic rocks.

Another approach which may help in determining the felsic/mafic percentage within the LC is constituted by the heat flux (HF), which in the Gran Sasso area varies from 50 to 60 mW/m$^2$ (Della Vedova et al., 2001). We can subtract from this value the amount of heat flux produced by U and Th (and K) contained within the SC (4.5 ± 1.6 mW/m$^2$) and the UC (18 ± 8.1 mW/m$^2$), using the abundances and the thickness already calculated and the atomic elemental heat production obtained by Fiorentini et al. (2005). The Upper Mantle temperature at the Moho depth can be determined based on thermo-barometric estimates of two mantle xenoliths suites (Veneto Volcanic Province and Iblei, Beccaluva et al., 2005). A temperature of 980°C can be estimated at 35 km (Moho depth in Central Italy varies between 25 and 35 km; Finetti et al., 2005b; see also Fig. 3), while 1100°C are measured at a depth of about 55 km, for a temperature gradient of ca. 6°C/km. According to these estimates and assuming a pure conductive heat flow (k = 4W/mK, Clauser and Huenges, 1995) the Upper Mantle can contribute to about 24 mW/m$^2$ to the whole heat flow (Boraso, 2008). Similar HF estimates (25 ± 5 mW/m$^2$) were determined for Central Italy by Cermak (1982), Hurtig and Stromeyer (1985) and Yegorova et al. (1997). Considering the central values, the HF of the LC can be estimated to 7.5 mW/m$^2$, which, according to the composition of the felsic and mafic reservoirs would require a completely felsic LC. Geological evidence however indicates that mafic rocks increase with depth, thus they cannot be less than 25%.

Taking into account the above reported geophysical and geochemical evidence the percentage of mafic rocks can be finally estimated to be:
$$m = 0.60 \pm 0.40$$
This result is in the felsic/mafic ratio ranges proposed by Wedephol (1995), Rudnick & Fountain (1995) and Gao et al (1998). Using this percentage the U and Th abundances in the whole LC can be obtained: $\alpha(U)=(0.3 \pm 0.3)$ ppm and $\alpha(Th)=(2.6 \pm 3.7)$ ppm (Table 11).

As in the case of UC, the ranges of published values for U and Th abundances in the LC, shown in the last line of Table 11 (Rudnick and Gao, 2003), essentially overlap with our estimate.

In conclusion, our study of the crust in central Italy returns U and Th concentrations that are similar to global averages of Upper and Lower Crust, but gives lower concentrations for the Sedimentary Cover, due to the high proportion of carbonates in the Gran Sasso area.

## 7. THE REFINED REFERENCE MODEL FOR THE GEO-NEUTRINO SIGNAL AT GRAN SASSO

At this stage, we have all the ingredients that are needed for a new estimate of the signal rate of geo-neutrinos from U and Th decay chains at the LNGS.

## 7.1 – The Regional Contribution

For the central tile, we use the 3D model developed above, which distinguishes eight reservoirs, organized as follow: four for the SC (a, b1, b2 and b3), two for UC (felsic and mafic) and two for LC (felsic and mafic). For each of these reservoirs, we adopt the U and Th abundances derived in this paper (Tables 8, 10 and 11), which are based on our own measurements. The antineutrino signals have been calculated in the same way as in Mantovani et al. (2004), by using these new inputs (Table 4, columns labeled as RRM).

For the rest of the regional area, we use the model developed in section 5.2, which distinguishes lower crust, upper crust and sediments, treated as a single and homogeneous layer. We adopt again our results for U and Th abundances. For the overlying sedimentary rocks, we assume U and Th abundances to be given by the weighted average performed according to the lithology in the Gran Sasso area, where U- and Th-poor carbonates account for some ¾ by mass of the whole sediments (see below for the consequences of this assumption).

The regional contribution from U+Th is 5 TNU lower than that of the RM (10.02 TNU in RRM, 15.59 TNU in RM; Table 4). The main reason for this difference is due to the treatment of the sediment layer in the central tile. Although, in this area the average Moho depth is close to the value found in Mantovani et al. (2004), the presence of a thick (some 13 km near Gran Sasso) sedimentary deposit composed primarily of U- and Th-poor carbonates essentially reduces the contribution to the signal. In contrast, assumptions about U and Th in the sediments for rest of the regional area have little impact on the estimated signal: if we used the world average abundances as for RM the Th+U signal would increase by only 0.5 TNU (from 0.34 TNU as in Table 4 to 0.84 TNU, this last value representing a simulation is not reported in Table 4).

## 7.2 – Rest of the Earth

Using the same geological framework as in the RM, we have updated the U and Th abundances in the different reservoirs, in accordance with recent reviews (see Table 3).

For the Upper and Middle Crust of the rest of the Earth, we adopt the values recommended in Rudnick and Gao (2003), which result from a detailed reanalysis of values presented in the literature and incorporating 1σ uncertainties. For the Lower Crust, values in the literature encompass a large interval, corresponding to different assumptions about the relative content of mafic/felsic rocks. We adopt here a mean value together with an uncertainty indicative of the spread of published values.

For sediments, we follow Plank and Langmuir (1998), as in the RM. For the UM, we follow Fogli (2006), who used an average between the results of Workman and Hart (2005) and of Salter and Stracke (2004). Concerning the BSE, we adopt the value provided in Palme and O'Neill (2003), in order to determine U and Th abundances in the lower mantle from mass balance.

In this way, the contribution form the rest of the Earth from U+Th is estimated as 26.1 TNU, which is very close to the value of 25.4 found with RM.

## 7.3 – The Predicted Signals in the RRM and Their Uncertainties

By summing the contribution of the different volumes and reservoirs we obtain the values shown in the last line of Table 4, which are the final estimates of our Refined Reference Model for the total geo-neutrino signals observed at LNGS,

$$S(U) = 28.7 \quad TNU \tag{7.3.1}$$

$$S(Th) = 7.5 \quad TNU \tag{7.3.2}$$

In total, S(U+Th) = 36.2 TNU, some 4 TNU smaller than the value of 40.5 (Table 4) predicted in the RM of Mantovani et al. (2004).

The assessment of the uncertainty of our prediction is a monumental task, since it results from several different sources of uncertainty, both statistical and systematic, which have to be combined together taking into account the possibility of correlations.

With the aim of obtaining an estimate of the uncertainty, we have propagated to the signals contributed from each reservoir r the uncertainties of the elemental abundances ar in that reservoir, (see Table 12 and Appendix 2). For each element, abundances in different reservoirs, except for the Lower Mantle (LM), are assumed to be affected by independent uncertainties. Uncertainties on the BSE abundances (which are the input for computing the LM abundances) are assumed to be independent from the other ones.

The resulting uncertainties form the regional area, $\Delta S_{reg}(U) = 3.1$ TNU and $\Delta S_{reg}(Th) = 0.9$ TNU are comparable to those from the rest of the Earth $\Delta S_{rest}(U) = 2.3$ TNU and $\Delta S_{rest}(Th) = 0.6$ TNU.

Uncertainties from the regional area and from the rest of the world can be considered as independent, and thus combined in quadrature in order to obtain an estimate of the global uncertainty:

$$\Delta S(U) = 3.9 \quad TNU \tag{7.3.3}$$
$$\Delta S(Th) = 1.0 \quad TNU \tag{7.3.4}$$

One has to observe that information on U and Th abundances are generally (at least partially) correlated, within each layer as well as for the BSE model: often, the abundance of one element is deduced from that of the other, assuming that the abundances ratio is better known. This is the case, for example, when rescaling the CI abundances in order to obtain the BSE estimate. Also, when considering the felsic/mafic rocks ratio in the crust, one introduces uncertainties that move U and Th abundances in the same direction. All these are positive correlations. Conservatively, we shall assume that the errors affecting the U and Th signal are fully positively correlated, i.e.:

$$\Delta S(U+Th) = \Delta S(U) + \Delta S(Th) = 4.9 \quad TNU \tag{7.3.5}$$

In conclusion, the uncertainty of our prediction on the U+Th signal is about 13%.

## 8. CONCLUDING REMARKS

The regional contribution to the geo-neutrino signal at Gran Sasso has been estimated based on a detailed geological, geochemical and geophysical study of the region; this is a necessary starting point if one wants to extract from the total signal the part that carries information on the global properties of the Earth.

A 3D model has been developed on an area 2° x 2° of latitude and longitude, centred on the LNGS and spanning four sedimentary layers, Upper Crust and Lower Crust. For the rest of the regional area, a simpler 3D model has been built, distinguishing three reservoirs only: sediments, Upper Crust and Lower Crust.

With the aim of assessing the U and Th content of rocks in the sediment layer in the Gran Sasso area and to verify their abundances in the Upper and Lower Crust presented in the literature, we have collected several samples from the Sedimentary Cover around Gran Sasso and from the crystalline basement in Northern Italy. We have measured U and Th abundances in these samples by using ICP-MS as well as scintillation (NaI) methods. The results were used to obtain our own estimates of abundances in the different layers of the region.

We have thus determined the contribution to the geo-neutrino signal originating from the region. When summed with the calculations for the rest of the world based on Mantovani et al. (2004) and using updated global abundances, we obtain the predictions of the Refined Reference Model (RRM, Table 4). The results,

$$S(U) = (28.7 \pm 3.9) \quad TNU \tag{8.1}$$

$$S(Th) = (7.5 \pm 1.0) \quad TNU \tag{8.2}$$

confirms, within errors, the values of the previous reference model (RM) developed in Mantovani et al. (2004).

With respect to the RM which calculated a total (Th+U) geo-neutrino signal of 40.5 (Table 4), the reduction is essentially due to the reduced content of U and Th in the thick sedimentary layer in the region where the detector is located (see section 5.1).

The interpretation of geoneutrino results requires detailed geological and geochemical study of the area around the detector. Such studies will be needed as the new geoneutrino detectors at SNO+, (Sudbury, Canada), LENA (Pyhasalmi, Finland) and Hanohano (Hawaii, USA) begin to acquire data. With appropriate modeling of the data from these detectors, the geo-neutrino signal attributable to the U and Th distribution in the Earth's mantle and even beyond can be identified. Further applications of geo-neutrinos and their implication for Earth Sciences have been recently discussed at the 4th Neutrino Geoscience (October, 6-7 2010, LNGS. http://geoscience.lngs.infn.it/).

**ACKNOWLEDGMENTS**

We are grateful to L. Beccaluva, E. Bellotti, C. Bonadiman, B. Della Vedova, L. Carmignani, A. Ianni, J. Majorowicz, W.F. Mc Donough, R. Rudnick and F. Siena for constructive discussions and useful comments. An earlier version of the manuscript was greatly improved thanks to the work of G. Herzog, M. Chen, A. Montanari and an anonymous referee. We wish to acknowledge the contribution which Nicola Ferrari gave to this work with his usual dedication, passion and intelligence. Sadly, an accident occurred on a mountain has taken his life, and has taken away from us an excellent colleague and a splendid friend. This work was partially supported by MIUR (Ministero dell'Istruzione, dell'Università e della Ricerca) under MIUR-PRIN-2006 project "Astroparticle physics" and by University of Ferrara under FAR2006 project "I geoneutrini come strumento di indagine geochimica del mantello terrestre".

# APPENDIX 1

# ERROR PROPAGATION: FROM ELEMENTAL ABUNDANCES TO GEO-NEUTRINO SIGNALS

**a) Regional Contribution**

With the aim of obtaining an indication of the uncertainty on the regional contribution to the signal, we have propagated to the signals $S_r(U)$ and $S_r(Th)$ from each reservoirs $r$ the uncertainties of the elemental abundances $a_r$ in each reservoirs, with the following criteria:

i. uncertainties on the CT are fully correlated with those of the rest of the regional area, as based on the same measurements and on the same hypothesis. We thus assume the relative error to be the same as that of CT and consider the whole regional area as a single block, subdivided into three reservoirs (Sediments, UC and LC)

ii. in each reservoir the contributed signal is proportional to the elemental abundance, $S_r = \alpha_r a_r$, so that the contributed error is

$$\Delta S_r = \alpha_r \Delta a_r$$

where for simplicity an index specifying the element (U or Th) is understood.

iii. uncertainties on the contribution of each reservoir are considered as independent of each other, as they derive from dispersions among the measurements of physically different sample sets; they will thus be combined in quadrature:

$$\Delta S_{reg} = \sqrt{\sum_r \Delta S_r^2}$$

For both U and Th, the contributed $\Delta S_r = \alpha_r \Delta a_r$ and the resulting $\Delta S_{reg}$ are calculated in table 12.

**b) Contribution from the Rest of the Earth**

Concerning uncertainties from the rest of the Earth, we proceed along similar lines, keeping into account the correlations imposed by the BSE mass constraint:

i. for each element (U or Th), we consider the abundance $a_{BSE}$ BSE and the abundances $a_r$ of all reservoirs but the LM as independent of each other.

ii. we remind that the abundance in the lower mantle $a_{LM}$ has been fixed by the BSE mass constraint, which gives:

$$a_{LM} = a_{BSE}\left(\frac{M_{BSE}}{M_{LM}}\right) - \sum_r a_r \left(\frac{M_r}{M_{LM}}\right)$$

where $M_r$ are the masses of the reservoirs and the sum contains all reservoirs but LM (reservoirs in the regional area can be neglected, since their masses are negligible on a global scale). The contributed signal form the rest of the Earth,

$$S_{rest} = \sum_r \alpha_r a_r + \alpha_{LM} a_{LM}$$

can thus be written in term of independent abundances as:

$$S_{rest} = \sum_r \left[\alpha_r - \alpha_{LM}\left(\frac{M_r}{M_{LM}}\right)\right] a_r + \alpha_{LM}\left(\frac{M_{BSE}}{M_{LM}}\right) a_{BSE}$$

iii. In this way, uncertainties from abundances can be added in quadrature:

$$\Delta S_{rest} = \sqrt{\sum_r \beta_r^2 \Delta a_r^2 + \beta_{BSE}^2 a_{BSE}^2}$$

where:

$$\beta_r = \alpha_r - \alpha_{LM}\left(\frac{M_r}{M_{LM}}\right)$$

$$\beta_{BSE} = \alpha_{LM}\left(\frac{M_{BSE}}{M_{LM}}\right)$$

For both U and Th, the contributed $\Delta S_r = \beta_r \Delta a_r$ and the resulting $\Delta S_{region}$ are calculated in table 12.

# APPENDIX 2

# GAMMA SPECTROMETRY

The samples have been measured through gamma spectrometry by means of a 3'x3' NaI crystal from ORTEC, installed in an underground building inside Hall A of the LNGS (Arpesella et al., 1996). This unique location guarantees a reduction of the cosmic ray flux by a factor $10^6$. The detector was positioned horizontally and enclosed in a lead shielding providing a coverage of at least 15 cm on all sides, in order to shield against the residual natural radioactivity of the environment.

The intrinsic background of the set-up has been routinely measured for a total time of 18.5 days. No significant variations from one background measurement to another have been observed. The global gain of the standard electronic chain (pre-amplifier, amplifier, ADC and Ortec-MAESTRO acquisition system) is also stable, allowing the addition of different energy spectra, without need of rebinning. The energy resolution of the detector is about 5% at the energy of the $^{40}$K line and varies with energy according to $E^{-1/2}$.

The main contributions to the background are due to cosmic rays, radioactivity from the detector and shielding materials together with X-rays, originated in the interactions of radiation within the lead shielding. Gamma peaks from the radionuclides $^{40}$K (1460.8 keV), $^{208}$Tl (2614.5 keV) and $^{214}$Bi (609.3 keV and 1764.5 keV) are clearly visible in the background spectrum; their contribution must be subtracted from the measured spectrum of the sample (see below).

The powder samples were kept in cylindrical plastic boxes of approximately 50 cm$^3$ volume, the rock samples were kept in similar boxes of larger volume, according to their sizes. The sample boxes were allocated inside especially designed polyethylene holders and positioned in front of the detector, in contact with its surface. The polyethylene holder keeps the sample box coaxial with the detector and at the same time minimizes the volume of air around the sample. For each sample, several spectra have been acquired in sequence, each one lasting $5 \cdot 10^4$ s. The number of runs has been chosen according to the counting rate of the sample, in order to have enough statistics in the relevant gamma peaks. The total counting times range from $1.7 \cdot 10^5$ s to $6.8 \cdot 10^5$ s. Typical counting rates are in the range from 0.7 to 2.7 cps in the energy interval (0.5 - 3.0) MeV.

The radionuclides of interest for the present work give the following contributions to the measured spectrum:
- a single line at 1460.8 keV and its Compton tail for $^{40}$K
- many lines, coming from different nuclides of their natural decay chains for $^{232}$Th, $^{238}$U and $^{235}$U.

It must be noted that only some of the elements of the total decay chain of U and Th contribute with detectable lines. In the $^{232}$Th chain, measurable lines are originated from $^{228}$Ac, $^{224}$Ra, $^{212}$Pb, $^{212}$Bi and $^{208}$Tl. Therefore, this measurement provides an equivalent Th content (Th$_{eq}$), which is equal to the real Th content under the assumption that secular equilibrium is respected. This is, for natural samples, a very likely assumption. In the case of $^{235}$U our instrumentation does not allow a quantitative evaluation of its activity, owing to the low isotopic abundance of this nuclide (0.7%), the low energy of the lines and the insufficient energy resolution of NaI. For $^{238}$U, the detectable lines are originated from $^{226}$Ra, $^{214}$Pb and $^{214}$Bi. We provide an equivalent U content (U$_{eq}$) from the detection of Bi and Pb lines. As in the case of Th, the U$_{eq}$ is equal to the content of U only if secular equilibrium is respected, which often is not the case.

For the determination of the Thorium content in the sample we use the strongest gamma line from $^{208}$Tl at 2614.5 keV (branching ratio 99%). This line, given the high gamma energy, lies in a region where the background from other radionuclides in the sample is essentially zero. For $^{238}$U we use the gamma line at 1764.5 keV from $^{214}$Bi. This is not the strongest line of the decay chain (branching ratio 15.4%) but it has the advantage to be isolated from other gamma lines and this makes easier to determine its intensity[2].

The counts in the relevant gamma peaks are obtained by fitting the experimental spectrum with a gaussian function overimposed on a linear background. To convert the number of counts into activity and then into concentration of the radionuclides of interest we have used the following formula:

$$c = \frac{C_s - C_b}{\varepsilon T m A}$$

where $C_s$ and $C_b$ are the counts in the gamma peak respectively in the sample and in the background spectrum, $T$ is the measuring time, $m$ is the mass of the sample, $\varepsilon$ is the counting efficiency and $A$ is the specific activity of the radionuclide[3].

In order to evaluate the counting efficiency of our detector we have used a Monte Carlo simulation program based on the Geant4 code (Agostinelli et al., 2003), widely used in the fields of highenergy, astroparticle and underground physics. It allows to generate primary particles ($e^{\pm}$, g-rays, ions, etc.), propagate them inside a given set-up and reconstruct the energy spectrum deposited inside the sensitive volume (the NaI crystal in our case). The geometry and the materials of the experimental set-up (sample, detector, shielding) must be defined by the user, while the accurate description of the physical process involved (radioactive decay, passage of particles through matter, energy deposition) and the properties of materials are provided by the code.

For the powder samples, which had all the same geometry, three spectra (one for each of the above listed nuclides) have been simulated assuming the nuclide to be uniformly distributed within the volume of the sample. For the rock samples, which had different shapes, we have used in the simulation code a parallelepiped shape with dimensions as close as possible to the real dimensions of the rock sample. This is an approximation which of course introduces a further inaccuracy in the simulation procedure (see below). The code generates directly the decay of the nuclide instead of the single gammas so the various branching ratios are correctly taken into account. The statistics of each simulation (i.e. the number of simulated decays for each nuclide) corresponds to $2 \cdot 10^6$ events for $^{214}$Bi and $^{208}$Tl and $5 \cdot 10^6$ events for $^{40}$K, given its low gamma yield. The efficiency is calculated simply as the ratio of the counts in the simulated peak and the total number of simulated events.

We have chosen to measure one of the first samples with a HPGe detector at LNGS, in order to have an independent confirmation of the reliability of the NaI measurements. The results of the two detectors show a good agreement. Moreover, from the observation of the $^{234m}$Pa gamma line at 1001 keV, we see no deviation from the secular equilibrium in the $^{238}$U decay chain.

For many of the measured samples the activity of the radio-nuclides of interest is well above the sensitivity of our detector but there are some cases where the net contribution of the sample to the measured spectrum is not statistically significant or is below the so-called limit of detection of the instrument. In order to treat correctly these cases we have referred to the procedure described in (Currie, 1968) and (Hurtgen, 2000). For each gamma peak in the measured spectrum we have calculated a decision-threshold ($L_c$) and a detection limit ($L_d$) defined in the following way:

$$L_c = 1.645\sqrt{2 C_b T_m / T_b}$$

$$L_d = 2 L_c$$

where $T_m$ and $T_b$ are the durations of the measurement and the background respectively and $k = 1.645$ is the coverage factor[4]. If the net signal in the gamma peak, defined as $C_{net} = (C_s - C_b)$ is higher than the detection limit, this means that there is a significant contribution of the sample and we can quote the result calculated following the above described procedure. If this is not the case the conclusion is that there is no significant contribution from the sample and we can only quote an upper limit on the contamination, according to the following prescriptions:

- if $C_{net} < L_c$ we quote the upper limit $C_{net} < L_d$;
- if $L_c < C_{net} < L_d$ we quote the upper limit $C_{net} < C_{net} + L_d$.

---

[2] This is not true for other gamma lines from $^{214}$Bi that, given the low energy resolution of the NaI detector, are always over imposed on lines from other elements, e.g. $^{208}$Tl.
[3] From IAEA report [IAEA, 2003].

The errors of the measurements reported in Tables 5 and 6a, 6b depends on the accuracy in calculating the following quantities:

- number of counts in the gamma peak;
- number of counts in the background peak;
- counting efficiency.

For the first two we rely on the error of the gaussian fit that we use to extract the number of counts in the peak. The uncertainty on the background counts are higher due to the lower statistics; this affects the result with a 5-7% error that we cannot avoid or reduce for the present experimental setup. The error in the determination of the counting efficiency depends on the accuracy of the Monte Carlo simulation which in turns is related to the accuracy of the GEANT4 code itself but mainly to the uncertainties in our description of the experimental setup which represents the input of the simulation program (detector and sample geometry, composition and density of the sample). In order to quantify this accuracy we have measured directly the detection efficiency at the $^{40}$K line using a KCl sample, whose content in K (52.44%) is known at the 0.7% level. The measured and simulated efficiencies agree within 2%.

As mentioned above, the efficiency depends on the density of the sample. The simulations have been repeated for 6 different values of the sample mean density, ranging from 1 to 1.8 g/cm$^3$, to take into account self-absorption in the sample. From 1 g/cm$^3$ to 1.4 g/cm$^3$ we have not observed significant variations of the efficiencies at the relevant energies. This is consistent with the fact that we are observing high-energy gamma peaks. Above 1.4 g/cm$^3$ we observe a decrease of the efficiency of about 3-4% by increasing the density of 0.2 g/cm$^3$. We assume a further contribution to the inaccuracy of 4% to account for the fact that we estimate the sample density with the ratio of the mass over the volume of the sample box and this is not really true if the sample box is not completely full.

The overall accuracy attributed to the determination of the counting efficiency can be then quantified in 5% for the powder samples and 10% for the rock samples where, as mentioned above, the inaccuracies in the description of the sample geometry must be taken into account, given the complicated sample shape.

---

[4] For this value of $k$, we are 95% certain that, if $L_c$ is exceeded, a net signal is really present.

TABLE 1 - The main properties of geo-neutrinos: for each decay chain Q is the Q-value, $\tau_{1/2}$ the half life of the parent nucleus, $E_{max}$ the maximal antineutrino energy, $Q_{eff} = Q - \langle E_{(\nu,\bar{\nu})} \rangle$, $\varepsilon_H$ and $\varepsilon_\nu$ are the heat and antineutrino production rate per unit mass and natural isotopic composition (see also Fiorentini et al., 2007)[5].

| Decay | $Q$ [MeV] | $T_{1/2}$ [$10^9$ yr] | $E_{max}$ [MeV] | $Q_{eff}$ [MeV] | $\varepsilon_H$ [W kg$^{-1}$] | $\varepsilon_{\bar{\nu}}$ [kg$^{-1}$ s$^{-1}$] |
|---|---|---|---|---|---|---|
| $^{238}$U -> $^{206}$Pb + 8 $^4$He + 6e + 6$\bar{\nu}$ | 51.7 | 4.47 | 3.26 | 47.7 | 0.95 10$^{-4}$ | 7.41 10$^7$ |
| $^{232}$Th -> $^{208}$Pb + 6 $^4$He + 4e + 4$\bar{\nu}$ | 42.7 | 14.0 | 2.25 | 40.4 | 0.27 10$^{-4}$ | 1.62 10$^7$ |
| $^{40}$K -> $^{40}$Ca + e + $\bar{\nu}$ (89%) | 1.311 | 1.28 | 1.311 | 0.590 | 0.22 10$^{-4}$ | 2.71 10$^4$ |

TABLE 2 - Predicted signal rates of geo-neutrinos from U + Th at various locations. Rates are in TNU (see text). All calculations are normalized to a survival probability $\langle P_{ee} \rangle$ = 0.57. For Mantovani et al. (2004) the uncertainties are estimated as ($N_{high} - N_{low}$)/6, where N is the total number of geoevents: $N_{high}$ and $N_{low}$ is the maximal and minimal prediction respectively (see also Table 12 of Mantovani et al. (2004).

| Location | Mantovani et al. (2004) | Fogli et al. (2006) | Enomoto et al. (2007)[6] |
|---|---|---|---|
| Hawaii | 12.5 ± 3.6 | 13.4 ± 2.2 | 13.4 |
| Kamioka | 34.8 ± 5.9 | 31.6 ± 2.5 | 36.5 |
| Gran Sasso | 40.5 ± 6.5 | 40.5 ± 2.9 | 43.1 |
| Sudbury | 49.6 ± 7.3 | 47.9 ± 3.2 | 50.4 |
| Pyhasalmi | 52.4 ± 7.6 | 49.9 ± 3.4 | 52.4 |

TABLE 3 - U and Th mass abundances in the Earth's reservoirs. Values adopted for regional and global estimates in the reference model (RM) can be found in Mantovani et al. (2004), while those used in the present work (RRM) are detailed in the text (see paragraph 7.2). Units are µg/g (ppm) and ng/g (ppb).

| Reservoir | Units | $a$(U) RM | $a$(Th) RM | $a$(U) RRM regional | $a$(Th) RRM regional | $a$(U) RRM global | $a$(Th) RRM global |
|---|---|---|---|---|---|---|---|
| Sediments | ppm | 1.68 | 6.9 | 0.78 ± 0.20 | 1.98 ± 0.44 | 1.68 ± 0.18 | 6.91 ± 0.8 |
| Upper Crust | ppm | 2.5 | 9.8 | 2.2 ± 1.2 | 8.3 ± 4.9 | 2.7 ± 0.6 | 10.5 ± 1.0 |
| Middle Crust | ppm | 1.6 | 6.1 | | | 1.3 ± 0.4 | 6.5 ± 0.5 |
| Lower Crust | ppm | 0.62 | 3.7 | 0.29 ± 0.24 | 3.17 ± 3.48 | 0.6 ± 0.4 | 3.7 ± 2.4 |
| Oceanic Crust | ppm | 0.1 | 0.22 | | | 0.1 ± 0.03 | 0.22 ± 0.07 |
| Upper Mantle | ppb | 6.5 | 17.3 | | | 3.95 ± 1.2 | 10.8 ± 3.24 |
| BSE | ppb | 20 | 78 | | | 21.8 ± 3.3 | 83.4 ± 12.5 |
| Lower Mantle | ppb | 13.2 | 52.0 | | | 16.7 | 57.4 |

---

[5] The antineutrino rate for unit mass of $^{235}$U at natural isotopic abundance composition (0.0072) is an order of magnitude less than those of $^{238}$U and $^{232}$Th. The antineutrinos from $^{235}$U are below the threshold for inverse beta on free protons.
[6] From [Enomoto et al., 2007] and private communication from S. Enomoto.

TABLE 4 - The contribution of the different reservoirs and areas to the geo-neutrino signal at Gran Sasso, in TNU units, according to the Reference Model of Mantovani et al. (2004) (RM) and to the Refined Reference model (RRM) presented in this paper. Results are presented for the approximate detector position (42° N, 14° E) used in Mantovani et al., (2004) and for the more precise value (42° 27' N, 13° 34' E) found in Bellotti (1988). Rest of Regional area = Tiles 1 to 6 minus Central Tile.

| Detector Latitude and Longitude | | 42° 27' N 13° 34' E | | | 42° 27' N 13° 34' E | 42° N 14° E |
|---|---|---|---|---|---|---|
| Area and reservoir | | S(U) RRM | S(Th) RRM | S(U+Th) RRM | S(U+Th) RM | S(U+Th) RM |
| **a) Regional contribution** | | | | | | |
| Central Tile (CT) | Sediments | 2.33 | 0.37 | 2.70 | 0.53 | 1.75 |
| | UC | 3.76 | 0.92 | 4.68 | 7.59 | 6.25 |
| | MC | = | = | = | 3.09 | 2.77 |
| | LC | 0.22 | 0.16 | 0.38 | 1.08 | 0.98 |
| Rest of the regional area | Sediments | 0.29 | 0.05 | 0.34 | 0.29 | 0.32 |
| | UC | 1.35 | 0.33 | 1.68 | 1.52 | 1.56 |
| | MC | = | = | = | 1.02 | 1.12 |
| | LC | 0.14 | 0.10 | 0.24 | 0.47 | 0.51 |
| *Regional Contribution, total* | | *8.09* | *1.93* | *10.02* | *15.59* | *15.26* |
| **b) Rest of the Crust** | | | | | | |
| Sediments | | 0.85 | 0.25 | 1.10 | 1.10 | 0.97 |
| Upper Crust | | 6.64 | 1.72 | 8.36 | 7.76 | 7.80 |
| Middle Crust | | 3.43 | 1.14 | 4.57 | 5.30 | 5.25 |
| Lower Crust | | 1.49 | 0.61 | 2.10 | 2.17 | 2.14 |
| Oceanic Crust | | 0.08 | 0.01 | 0.09 | 0.09 | 0.09 |
| *Rest of the crust, total* | | *12.49* | *3.73* | *16.22* | *16.42* | *16.25* |
| **c) Mantle** | | | | | | |
| Upper Mantle | | 0.86 | 0.16 | 1.02 | 1.68 | 1.68 |
| Lower Mantle | | 7.24 | 1.65 | 8.89 | 7.32 | 7.32 |
| *Mantle, total* | | *8.10* | *1.81* | *9.91* | *9.0* | *9.0* |
| **d) Earth, total** | | **28.7** | **7.5** | **36.2** | **41.0** | **40.5** |

TABLE 5 - Sedimentary reservoirs. GPS coordinates of collected samples and their distance in km from LNGS are reported. a) Cenozoic terrigenous units (sandstones, siltites and clays), b1) Meso-Cenozoic basinal carbonate units (marly and shaly carbonates, sometimes with black shales), b2) Mesozoic carbonate units (limestones, dolomites and evaporites, with a negligible marl and clay content). The sedimentary formations to which samples belongs are reported in Table 7. U and Th abundances are in ppm. Uncertainties are the instrumental 1σ error. Measurement methods are indicated in parenthesis. Adopted values are in bold. Values in brackets are not considered (see text for explanations).

| N. | Sample Code | Latitude | Longitude | Distance [km] | Reservoir | U (ICP-MS) | $U_{eq}$ (NaI) | Th (ICP-MS) | $Th_{eq}$ (NaI) |
|---|---|---|---|---|---|---|---|---|---|
| 1 | 11TS | 42° 32' 52" | 13° 29' 00" | 13 | a | **2.64 ± 0.26** | 2.80 ± 0.60 | **10.19 ± 1.02** | 10.6 ± 1.7 |
| 2 | 12TS | 42° 33' 09" | 13° 28' 22" | 14 | a | **2.93 ± 0.29** | 2.00 ± 0.40 | **10.54 ± 1.05** | 8.3 ± 1.2 |
| 3 | 13TC | 42° 24' 28" | 13° 49' 36" | 17 | a | **1.60 ± 0.16** | 2.10 ± 0.80 | **5.53 ± 0.55** | 4.4 ± 1.0 |
| 4 | 14TC | 42° 29' 11" | 13° 43' 09" | 13 | a | **1.94 ± 0.19** | 1.30 ± 0.60 | **6.97 ± 0.70** | 6.6 ± 1.6 |
| 5 | 06MM | 42° 28' 57" | 13° 32' 33" | 4 | b1 | **0.49 ± 0.05** | 0.18 ± 0.12 | **1.32 ± 0.13** | 1.3 ± 0.3 |
| 6 | 07MM | 42° 28' 39" | 13° 32' 21" | 4 | b1 | **0.94 ± 0.09** | 0.60 ± 0.30 | **0.32 ± 0.03** | 0.38 ± 0.32 |
| 7 | 08MM | 42° 28' 27" | 13° 31' 51" | 4 | b1 | (3.43 ± 0.34) | (1.15 ± 0.22) | **0.21 ± 0.02** | 1.0 ± 0.3 |
| 8 | 09OR | 42° 25' 10" | 13° 22' 38" | 16 | b1 | **0.95 ± 0.09** | 0.35 ± 0.31 | **4.23 ± 0.42** | 2.8 ± 0.8 |
| 9 | 10OR | 42° 27' 22" | 13° 21' 14" | 17 | b1 | **2.73 ± 0.27** | 2.20 ± 0.40 | **0.23 ± 0.02** | < 0.1 |
| 10 | CPN1 | 42° 26' 50" | 13° 26' 13" | 12 | b1 | **3.94 ± 0.39** | 2.50 ± 0.30 | **3.37 ± 0.34** | 1.1 ± 0.4 |
| 15 | MB8 | 42° 44' 45" | 13° 37' 17" | 33 | b1 | **0.38 ± 0.04** | | **0.96 ± 0.10** | |
| 16 | MB14 | 42° 44' 35" | 13° 35' 38" | 33 | b1 | **0.12 ± 0.01** | | **0.11 ± 0.01** | |
| 17 | MB23 | 43° 21' 54" | 13° 02' 57" | 110 | b1 | **0.48 ± 0.05** | | **0.24 ± 0.02** | |
| 18 | MB22 | 43° 21' 44" | 13° 03' 10" | 110 | b1 | **0.30 ± 0.03** | | **0.91 ± 0.09** | |
| 19 | MB16 | 43° 32' 49" | 13° 38' 03" | 121 | b1 | **5.85 ± 0.58** | | **4.01 ± 0.40** | |
| 20 | MB34B | 42° 43' 45" | 13° 26' 40" | 32 | b1 | **2.44 ± 0.24** | | **2.07 ± 0.02** | |
| 11 | 02CM | 42° 18' 26" | 13° 41' 41" | 16 | b2 | **0.11 ± 0.01** | < 0.1 | **0.06 ± 0.01** | < 0.1 |
| 12 | 03CP | 42° 25' 11" | 13° 20' 58" | 18 | b2 | **0.58 ± 0.06** | 0.40 ± 0.18 | **0.03 ± 0.01** | < 0.1 |
| 13 | 04CS | 42° 27' 54" | 13° 32' 04" | 3 | b2 | **0.12 ± 0.01** | < 0.1 | **0.42 ± 0.04** | < 0.3 |
| 14 | 05CS | 42° 28° 28" | 13° 32' 26" | 3 | b2 | **0.39 ± 0.04** | 0.60 ± 0.30 | **0.31 ± 0.03** | 0.3 ± 0.4 |
| 21 | APN5 | 42° 17' 54" | 13° 52' 25" | 30 | b2 | **0.43 ± 0.04** | 0.60 ± 0.26 | **0.23 ± 0.02** | < 0.3 |
| 22 | GAN3 | 42° 17' 57" | 13° 51' 52" | 30 | b2 | **0.22 ± 0.02** | 0.37 ± 0.17 | **0.08 ± 0.01** | 0.4 ± 0.2 |
| 23 | MB5 | 42° 44' 51" | 13° 37' 28" | 33 | b2 | **0.60 ± 0.06** | | **0.17 ± 0.02** | |
| 24 | MB6 | 42° 44' 48" | 13° 37' 25" | 33 | b2 | **0.18 ± 0.02** | | **0.14 ± 0.01** | |
| 25 | MB9 | 42° 44' 44" | 13° 37' 10" | 33 | b2 | **0.77 ± 0.08** | | **0.02 ± 0.01** | |
| 26 | MB18 | 43° 23' 19" | 13° 02' 59" | 112 | b2 | **0.21 ± 0.02** | | **0.02 ± 0.01** | |
| 27 | MB19 | 43° 22' 03" | 13° 01' 58" | 111 | b2 | **0.20 ± 0.02** | | **0.07 ± 0.01** | |
| 28 | MB20 | 43° 21' 41" | 13° 02' 55" | 110 | b2 | **0.14 ± 0.01** | | **0.68 ± 0.07** | |

TABLE 6A – Upper crust reservoirs. UC samples are taken from Val Strona, Serie dei Laghi (label VS), Valsugana and Cima d'Asta-Caoria (numbered samples and label CA). U and Th abundances are in ppm. Uncertainties are the instrumental 1σ error. Measurement methods are indicated in parenthesis. Adopted values are in bold. Values in brackets are not considered (see text for explanations). /, not measured.

| N. | Sample Code | Composition | Lithology | U (ICP-MS) | U (NaI) | U (adopted) | Th (ICP-MS) | Th (NaI) | Th (adopted) |
|---|---|---|---|---|---|---|---|---|---|
| 1 | VS11 | mafic | anfibolite | 0.59 ± 0.06 | 0.4 ± 0.3 | **0.59 ± 0.06** | 0.43 ± 0.04 | <0.3 | **0.43 ± 0.04** |
| 2 | VS15 | mafic | anfibolite | 0.13 ± 0.01 | <0.10 | **0.13 ± 0.01** | 0.08 ± 0.01 | <0.3 | **0.08 ± 0.01** |
| 3 | VS7 | felsic | Migmatitic kinzigite | 1.92 ± 0.19 | / | **1.92 ± 0.2** | 11.7 ± 1.17 | / | **11.7 ± 1.2** |
| 4 | VS9 | felsic | kinzigite | 3.62 ± 0.36 | 3.1 ± 0.4 | **3.1 ± 0.4** | 18.0 ± 1.80 | 13.7 ± 1.20 | **15.0 ± 1.0** |
| 5 | VS10 | felsic | kinzigite | 2.40 ± 0.24 | 3.2 ± 0.5 | **3.2 ± 0.5** | 10.3 ± 1.03 | 10.3 ± 1.30 | **10.3 ± 0.8** |
| 6 | VS18 | felsic | kinzigite | 2.00 ± 0.20 | 2.9 ± 0.6 | **2.9 ± 0.6** | 9.71 ± 0.97 | 10.0 ± 1.10 | **9.8 ± 0.7** |
| 7 | VS19 | felsic | aplite | 1.68 ± 0.17 | 2.4 ± 0.5 | **2.4 ± 0.5** | 15.3 ± 1.53 | 13.5 ± 1.20 | **14.2 ± 0.9** |
| 8 | VS20 | felsic | granit | 3.05 ± 0.30 | 5.9 ± 0.7 | **5.9 ± 0.7** | 14.4 ± 1.44 | 15.6 ± 1.60 | **14.9 ± 1.1** |
| 9 | VS21 | felsic | granit | 3.24 ± 0.32 | 3.5 ± 0.6 | **3.5 ± 0.6** | 12.8 ± 1.28 | 13.8 ± 1.30 | **13.3 ± 0.9** |
| 10 | VS22 | felsic | schist | 1.82 ± 0.18 | 2.2 ± 0.5 | **2.2 ± 0.5** | 10.8 ± 1.08 | 10.0 ± 1.20 | **10.4 ± 0.8** |
| 11 | VS23 | Felsic | Micascisto | 2.03 ± 0.20 | 2.8 ± 0.6 | **2.8 ± 0.6** | 11.9 ± 1.19 | 9.4 ± 1.10 | **10.6 ± 0.8** |
| 12 | 14 | felsic | diorite | 1.44 ± 0.14 | 1.7 ± 0.3 | **1.7 ± 0.3** | 9.35 ± 0.94 | 7.6 ± 0.80 | **8.3 ± 0.6** |
| 13 | 16 | felsic | monzogranite | 2.05 ± 0.21 | 3.1 ± 0.5 | **3.1 ± 0.5** | 9.77 ± 0.98 | 9.1 ± 0.98 | **9.4 ± 0.7** |
| 14 | 26 | felsic | phyllades-schists | 1.30 ± 0.13 | 2.4 ± 0.4 | **2.4 ± 0.4** | 5.01 ± 0.50 | 5.4 ± 0.70 | **5.1 ± 0.4** |
| 15 | 27 | felsic | phyllades-schists | 1.78 ± 0.18 | 2.2 ± 0.5 | **2.2 ± 0.5** | 11.6 ± 1.16 | 11.0 ± 1.25 | **11.3 ± 0.9** |
| 16 | CA58 | felsic | phyllades-schists | 2.34 ± 0.23 | 2.6 ± 0.5 | **2.6 ± 0.5** | 12.5 ± 1.25 | 12.6 ± 1.20 | **12.6 ± 0.9** |
| 17 | CA63 | Felsic | phyllades-schists | 1.77 ± 0.18 | 3.3 ± 0.6 | **3.3 ± 0.6** | 9.65 ± 0.97 | 12.8 ± 1.40 | **10.7 ± 0.8** |
| 18 | CA65 | felsic | phyllades-schists | 1.61 ± 0.16 | 3.2 ± 0.5 | **3.2 ± 0.5** | 7.19 ± 0.72 | 14.4 ± 1.30 | (8.88 ± 0.63) |
| 19 | CA69 | felsic | phyllades-schists | 1.45 ± 0.15 | 2.2 ± 0.4 | **2.2 ± 0.4** | 12.3 ± 1.23 | 16.2 ± 1.40 | **14.0 ± 0.9** |
| 20 | CA74 | felsic | tonalite | 1.57 ± 0.16 | 2.4 ± 0.5 | **2.4 ± 0.5** | 6.71 ± 0.67 | 5.4 ± 0.70 | **6.1 ± 0.5** |

TABLE 6B – Lower crust reservoirs. LC samples are taken from Ivrea-Verbano Zone (Val Strona, Val Sessera and Val Sesia). U and Th abundances are in ppm. Uncertainties are the instrumental 1σ error. Measurement methods are indicated in parenthesis. Adopted values are in bold. Values in brackets are not considered (see text for explanations). /, not measured.

| N. | Sample Code | Composition | Lithology | U (ICP-MS) | U (NaI) | U (adopted) | Th (ICP-MS) | Th (NaI) | Th (adopted) |
|---|---|---|---|---|---|---|---|---|---|
| 1 | VS1 | mafic | amph-gabbro | 0.02 ± 0.01 | / | **0.02 ± 0.01** | <0.01 | / | **0.005 ± 0.005** |
| 2 | VS6 | mafic | diorite | 0.35 ± 0.03 | 1.04 ± 0.23 | **0.35 ± 0.03** | 0.84 ± 0.08 | 0.9 ± 0.33 | **0.84 ± 0.08** |
| 3 | VS8 | mafic | gabbro | 0.05 ± 0.01 | <0.1 | **0.05 ± 0.01** | 0.03 ± 0.01 | <0.3 | **0.03 ± 0.01** |
| 4 | VS13 | mafic | granulite | <0.01 | <0.1 | **0.01 ± 0.01** | 0.09 ± 0.01 | <0.3 | **0.09 ± 0.01** |
| 5 | VS2 | felsic | stronalite | 0.42 ± 0.04 | 0.9 ± 0.35 | **0.42 ± 0.04** | 5.46 ± 0.55 | 12.9 ± 1.3 | (6.59 ± 0.51) |
| 6 | VS3 | felsic | charnockite | 0.01 ± 0.01 | <0.1 | **0.01 ± 0.01** | 0.01 ± 0.01 | <0.3 | **0.01 ± 0.01** |
| 7 | VS5 | felsic | stronalite | 0.10 ± 0.01 | / | **0.10 ± 0.01** | 0.17 ± 0.02 | <0.3 | **0.17 ± 0.02** |
| 8 | VS12 | felsic | stronalite | 1.14 ± 0.11 | 1.3 ± 0.4 | **1.14 ± 0.11** | 11.50 ± 1.15 | 12.6 ± 1.30 | **12.0±0.9** |
| 9 | VS14 | Felsic | Stronalite | 0.70 ± 0.07 | 0.7 ± 0.3 | **0.70 ± 0.07** | 15.40 ± 1.54 | 10.7 ± 1.10 | **12.3±0.9** |

TABLE 7 – Lithological description, thickness (minimum and maximum, in meters), age, depositional environment and attribution to the various reservoirs for the main lithostratigraphic units of the Central Apennines. Data compiled from many Authors (Crescenti et al., 1969; Bernoulli, 1972, 2001; Parotto and Praturlon, 1975; Coltorti and Bosellini, 1980; Vai, 2001; Finetti et al., 2005a among others).

| Formations | Lithologies | min thick. | max thick. | Age | Depositional environment |
|---|---|---|---|---|---|
| Verrucano | Conglomerate and sandstones | 600 | 700 | Permian-Trias | continental |
| Anidriti di Burano | Anhydrites and dolostones, sometimes with breccia intervals | 1000 | 1700 | Norian-Rhaetian | sabka |
| Calcari bituminosi (Pietre Nere) | Limestone very dark in colour with some pyrite framboids | 0 | ? | Carnian-Norian | anoxic basin |
| Dolomie di Castelmanfrino | Dolostones, locally peloidal limestones, bedded to massive, with chert levels and nodules | 1030 | 1030 | Lias | inner platform and basin |
| Calcari a Rhaetavicula - M. Cetona | Limestone and marls passing to dolostones in the upper part | 30 | 150 ? | Rhaetian | inner platform |
| Calcare Massiccio | Skeletal limestone, lime mudstone, oolitic to peloidal limestone and stromatolitic intervals | 80 | 600 | Hettangian-Sinemurian | inner platform |
| Corniola | Lime mudstone with cherts nodules, thinly bedded, sometimes breccias and resediments | 150 | 230 | Sinemurian p.p.-Pliensbachian | basin |
| Marne del Serrone | Marls and clay with marly limestone, with a black shale | 20 | 60 | Toarcian | basin |
| Rosso Ammonitico | Marls and limestone with platform derived resediments | 0 | 45 | Toarcian | basin |
| Verde Ammonitico | Marls and limestone, pass laterally to Marne del Serrone | 0 | 60 | Toarcian | basin |
| Calcari e Marne a Posidonia | lime mudstone with skeletal beds | 20 | 50 | Aalenian - Bajocian/Bathonian | basin |
| Calcari ad Aptici and Calcari Diasprigni | limestones with frequent chert in levels and nodules decreasing upward | 50 | 140 | Callovian-Tithonian p.p. | basin |
| Bioclastici inferiori | Massive to thick beds of platform derived non-skeletal and skeletal resediments | 0 | 135 | Bajociano p.p. - Kimmeridgiano inf. | slope |
| Calcari ad Ellipsactinie | Massive bioclastic limestones rich in reef biota | 150 | 250 | Kimmeridgiano - Titonico p.p. | platform margin |
| Maiolica | Lime mudstone with cherts nodules, thinly bedded | 20 | 400 | Tithonian p.p. - Early Aptian | basin |
| Bioclastici superiori | lithoclastic breccias with thinnely bedded limestones and marly limestones | 0 | 60 | Albiano p.p. Cenomaniano p.p. | slope |
| Marne a Fucoidi | Marly limestone and marls, with some silicified intervals and thin black shales | 10 | 100 | Aptian p.p-Cenomanian p.p | basin |
| Scaglia Bianca | Lime mudstone with cherts nodules, thinly bedded | 0 | 45 | Late Albian - Early Turonian | basin |
| Scaglia Rossa | Lime mudstone with cherts nodules, thinly bedded | 200 | 400 | Early Turonian - Middle Eocene | basin |
| Orfento | Thick to massive bioclastic limestones | 50 | 200 | Campanian p.p. Maastrichtiano p.p. | proximal to distal ramp |
| Scaglia Cinerea | Marly limestone and marls, with some chert intervals | 60 | 190 | Late Eocene- Late Oligocene | basin |
| Glauconitico | lithoclastic limestones, marly limestone spongolithic levesl and chert in levels or nodules | 10 | 80 | Aquitanian-Burdigalian | ramp |
| Bisciaro | Marls and marly limestone with some volcaniclastic layers, sometimes with chert nodules | 15 | 150 | Aquitanian-Langhian | basin |
| Marne con Cerrogna | Marls, marly limestones and clayely marls, with calcarenitic and calcirudite intercalations | 60 | 300 | Langhian p.p.-Tortonian? | base-of-slope to basin |
| Marne a Orbulina | thinly bedded clayely marls | 0 | 150 | Tortonian-Messinian | foredeep |
| Laga | Turbiditic sandstones and marls alternations, with gypsarenites intercalated | 1400 | 4000 | Messinian | foredeep |
| Trave | Marls and dark clay interbedded | 40 | 120 | Messinian | shelf to shallow basin |
| Colombacci | marly clays with thinly bedded arenites in the lower part and conglomerates in the upper part | 50 | 250 | Messinian post-evaporitic | from basin to lacustrine |
| Schlier | marls and clayey marls with subordinated clayey limestones and resedimented levels | 400 | 800 | Burdigalian - Early Messinian | basin |
| Cellino | Turbiditic sandstones and marls alternations thinning and fining upward | 800 | 2500 | Early Pliocene | foredeep |
| Argille azzurre | silty clays, marly clays and clayely marls, with some siltstones, sandstones and conglomerates | 1000 | 3500 | Pliocene-Pleistocene p.p. | basin |
| Morrone di Pacentro | micritic limestones with intercalations of grainstone and packstone | 2000 | 3000 | Turonian-early Paleocene | inner platform |
| Monte Acquaviva | micritic limestones, calcarenites, calcirudites and rudist biostromes | 200 | 500 | Allenian-Albian | platform margin to base-of-slope |
| Terratta | bioclastic calcarenites and calcirudites, local biohermos, and micritic limestones | 500 | 2000 | Late Paleocene-Early Oligocene | proximal to distal ramp |
| Santo Spirito | fine to coarse grained calcarenites, marly limestones and chert in levels and nodules | 0 | 300 | Late Oligocene-Early Messinian | proximal to distal ramp |
| Bolognano | fine to coarse greined calcarenites, marly limestones | 0 | 200 | Late Oligocene-Early Messinian | proximal to distal ramp |
| Calcari a Briozoi e lilotamni | fine to coarse grained calcarenites, marly limestones | 0 | 200 | Late Oligocene-Early Messinian | proximal to distal ramp |

| Reservoir | Samples |
|---|---|
| b3 | APN5, GAN3 |
| b2 | CPN1 |
| b1 | |
| b2 | |
| b2 | |
| (b2) | |
| b2 | 03CP, MB18 |
| b1 | MB14, MB23 |
| b1 | |
| (b1) | |
| b1 | 09OR |
| (b1) | |
| b1 | MB22 |
| (b1) | |
| b2 | MB19, MB20 |
| (b1) | |
| b2 | 05CS, MB9 |
| b2 | 02CM |
| (b1) | |
| b2 | MB8, 10OR |
| b2 | MB5, MB6 |
| (b1) | |
| b2 | 04CS |
| (b2) | |
| b1 | 08MM |
| b1 | MB34b |
| b1 | 07MM |
| b1 | 06MM |
| a | 11TS, 12TS, 13TC, 14TC |
| b1 | MB16 |
| (a) | |
| (a) | |
| (a) | |
| (b2) | |
| (b2) | |
| (b2) | |
| (b2) | |
| (b2) | |
| (b2) | |

TABLE 8 – Calculated U and Th abundances for the whole sedimentary cover (SC) of the Central Tile. Approximate average densities are from Telford et al. (1990) and are in agreement with the densities assumed by Laske et al. (2001) for the rest of the world.

| Reservoir | | Density [Ton/m$^3$] | Volume [%] | Mass [%] | $a$(U) [ppm] | $a$(Th) [ppm] |
|---|---|---|---|---|---|---|
| a) Cenozoic terrigenous sediments | | 2.1 | 18.0 | 15.6 | 2.3 ± 0.6 | 8.3 ± 2.5 |
| b) Permo-Mesozoic carbonatic succession | b1) Meso-Cenozoic Basinal Carbonates | 2.3 | 2.0 | 1.8 | 1.7 ± 1.8 | 1.5 ± 1.6 |
| | b2) Mesozoic Carbonates | 2.5 | 74.6 | 76.8 | 0.3 ± 0.2 | 0.2 ± 0.2 |
| | b3) Permian clastic units | 2.6 | 5.4 | 5.8 | 2.2 ± 1.3 | 8.1 ± 4.9 |
| Mass weighted averages | | | | | 0.8 ± 0.2 | 2.0 ± 0.5 |
| Values used in the Reference Model | | | | | 1.7 | 6.9 |

TABLE 9 – Volume (in km$^3$ and %) and thickness (in km) of the four sedimentary reservoirs and of the UC and LC. Thickness are reported according to the present model (RRM) and the Reference Model (RM, Mantovani et al., 2004).

| | Volume (km$^3$) | Volume (%) | Thickness (km) | |
|---|---|---|---|---|
| | | | RM | RRM |
| a) Cenozoic terrigenous sediments | 83,589 | 6.8 | 0.5 | 13 |
| b1) Meso-Cenozoic Basinal Carbonates | 9,028 | 0.7 | | |
| b2) Mesozoic Carbonates | 345,684 | 28.1 | | |
| b3) Permian clastic units | 25,163 | 2.0 | | |
| Upper crust | 468,772 | 38.0 | 10 | 13 |
| Middle crust | / | / | 10 | / |
| Lower crust | 300,566 | 24.4 | 10.5 | 9 |
| **Total** | **1,232,802** | **100** | **31** | **35** |

TABLE 10 – U and Th abundances for the Upper Crust obtained from mass weighted average. Approximate average densities are from Telford et al. (1990) and are in agreement with the densities assumed by Laske et al. (2001) for the rest of the world. Uncertainties on the mass weighted average are calculated taking also into account the spread on the mafic/felsic ratios.

| Reservoir | | Density [Ton/m$^3$] | Volume [%] | Mass [%] | $a$(U) [ppm] | $a$(Th) [ppm] |
|---|---|---|---|---|---|---|
| Upper Crust | Mafic | 3.0 | 25 | 27 | 0.4 ± 0.3 | 0.3 ± 0.2 |
| | Felsic | 2.7 | 75 | 73 | 2.8 ± 0.9 | 11.0 ± 2.8 |
| Mass weighted average | | | | | 2.2 ± 1.3 | 8.1 ± 4.9 |
| Values used in the Reference Model | | | | | 2.5 | 9.8 |

TABLE 11 – U and Th abundances for the lower crust.

| Reservoir | | $a$(U) [ppm] | $a$(Th) [ppm] |
|---|---|---|---|
| Lower Crust | Mafic | 0.1 ± 0.2 | 0.2 ± 0.4 |
| | Felsic | 0.5 ± 0.5 | 6.1 ± 7.0 |
| Mass weighted average | | 0.3 ± 0.3 | 2.6 ± 3.7 |
| Values used in the Reference Model | | 0.6 | 3.7 |

TABLE 12– Estimated uncertainties on the geo-neutrino signal, in TNU (see Appendix 2).

| Area and reservoir | ΔS(U) RRM | ΔS(Th) RRM |
|---|---|---|
| **a) Regional contribution** | | |
| Sediments | 0.66 | 0.11 |
| Upper Crust | 3.02 | 0.76 |
| Middle Crust | | |
| Lower Crust | 0.36 | 0.37 |
| *Regional Contribution, total* | *3.11* | *0.85* |
| **b) Rest of the Earth** | | |
| Sediments | 0.06 | 0.02 |
| Upper Crust | 0.86 | 0.10 |
| Middle Crust | 0.64 | 0.05 |
| Lower Crust | 0.61 | 0.24 |
| Oceanic Crust | 0.003 | 0.001 |
| Upper Mantle | 0.07 | 0.014 |
| BSE | 1.96 | 0.49 |
| *Rest of the Earth, total* | *2.32* | *0.56* |
| **c) Earth, total** | **3.9** | **1.0** |

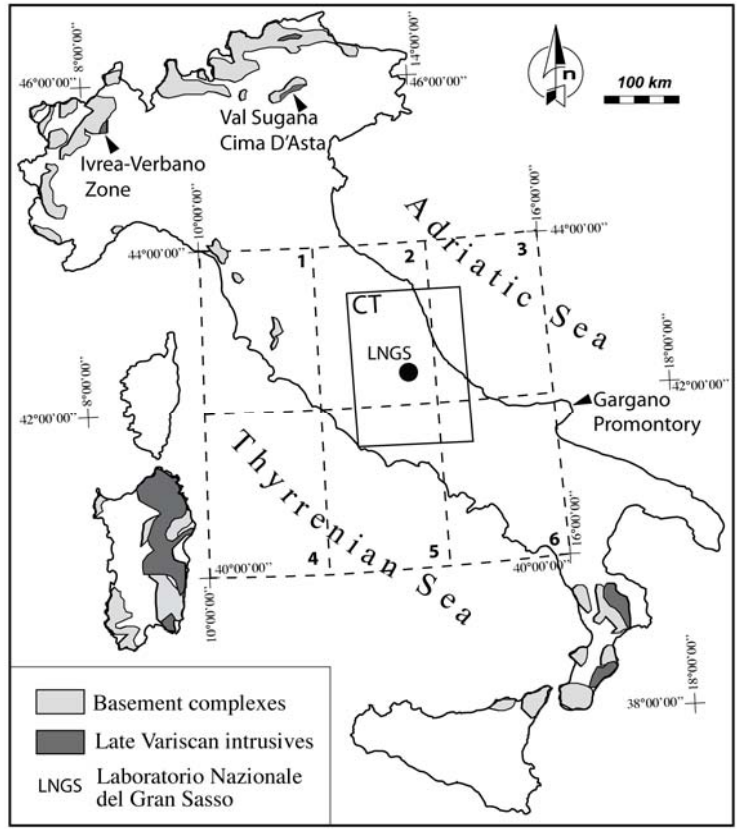

Fig. 1 – Schematic map of the main area where basement complexes and Late Variscan intrusive bodies crop out in Italy (modified after Boriani et al., 2003). The central tile (CT) of the 2°x2° centered at the Gran Sasso laboratory (LNGS; 42°27' N, 13°34 'E) and the six tiles which provide the "regional contribution" to the geoneutrino flux are also reported.

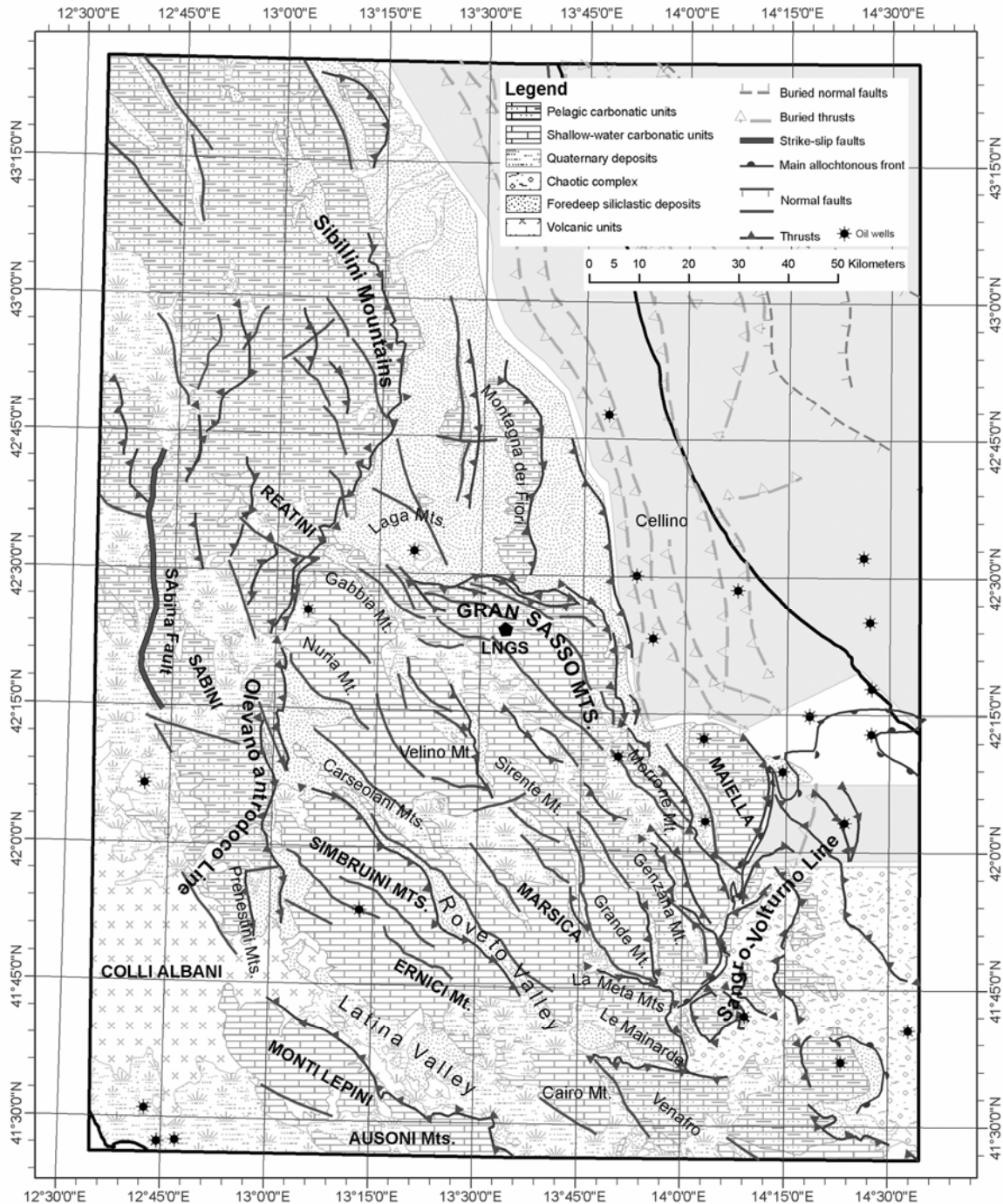

Fig. 2 – Geological map of the central tile (CT). Data are stored in a GIS (Global Information System) frame and, together with the CROP (CROsta PRofonda, i.e. deep crust) seismic profiles of Finetti et al. (2005a), are used to reconstruct the 3D geological model. This area enclosed the whole central Apennines and some part of the northern Apennines (west of the Olevano-Antrodoco Line, previously known as Ancona-Anzio Line) and southern Apennines (east of the Sangro-Volturno Line, previously known as Ortona-Roccamonfina Line) (mainly from Vezzani and Ghisetti, 1998 and Finetti et al., 2005a, among many others). The central part is mainly constituted by Mesozoic carbonate platform, passing northward to Tertiary siliciclastic foredeep deposits (Laga Fm.). The northern Apennines are characterized by carbonatic basinal facies while the southern Apennines by argillaceous chaotic complexes. Asterisks represent the most representative and deep wells used in the model.

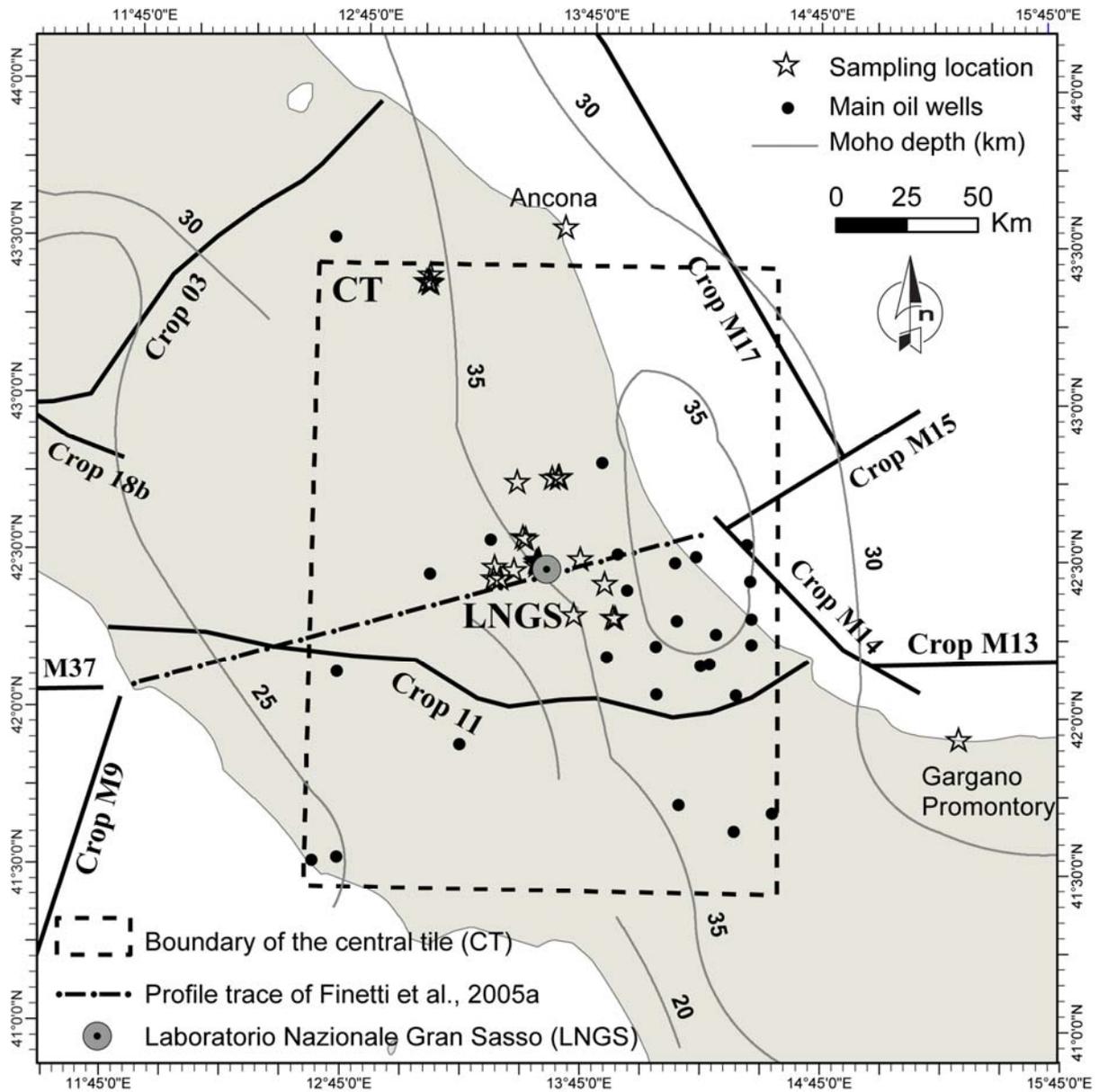

Fig. 3 – The main inputs used for building the 3D model of the Central Tile. The dot-dashed line (RLSS) corresponds to the Reconstructed Lithospheric Seismological Section of (Finetti, 2005a). EPC11 denotes the eastern part of CROP 11, used for comparison of different investigations. Also the locations of the Sedimentary Cover samples used in this work are shown.

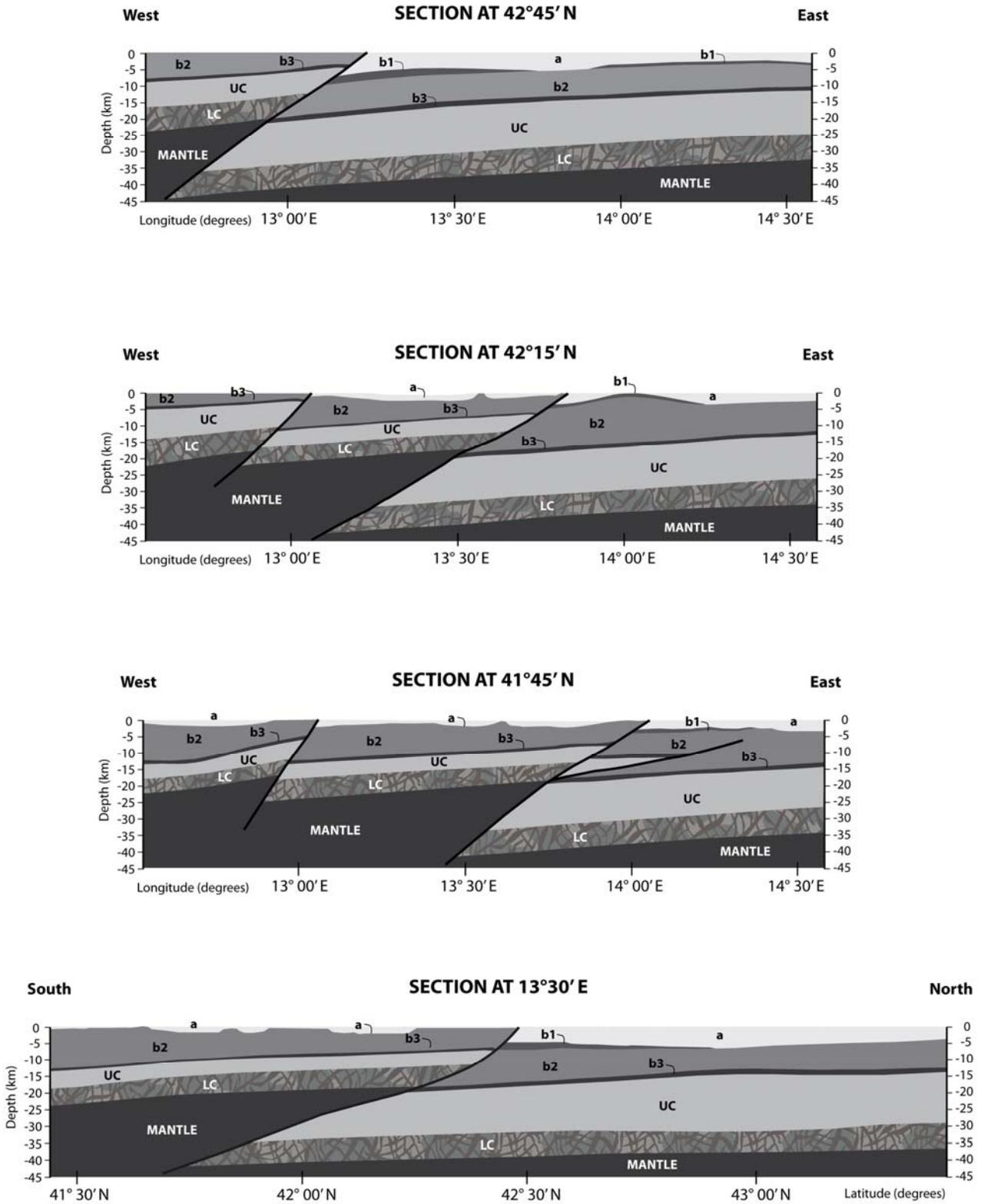

Fig. 4 – Schematic E-W and N-S cross sections of the main reservoirs in the central tile reconstructed on the basis of the 3D model. a, b1, b2 and b3 refer to Cenozoic terrigenous sediments, Meso-Cenozoic Basinal Carbonates, Mesozoic Carbonates and Permian clastic units; UC and LC to Upper crust and Lower Crust respectively.

a) Sediments

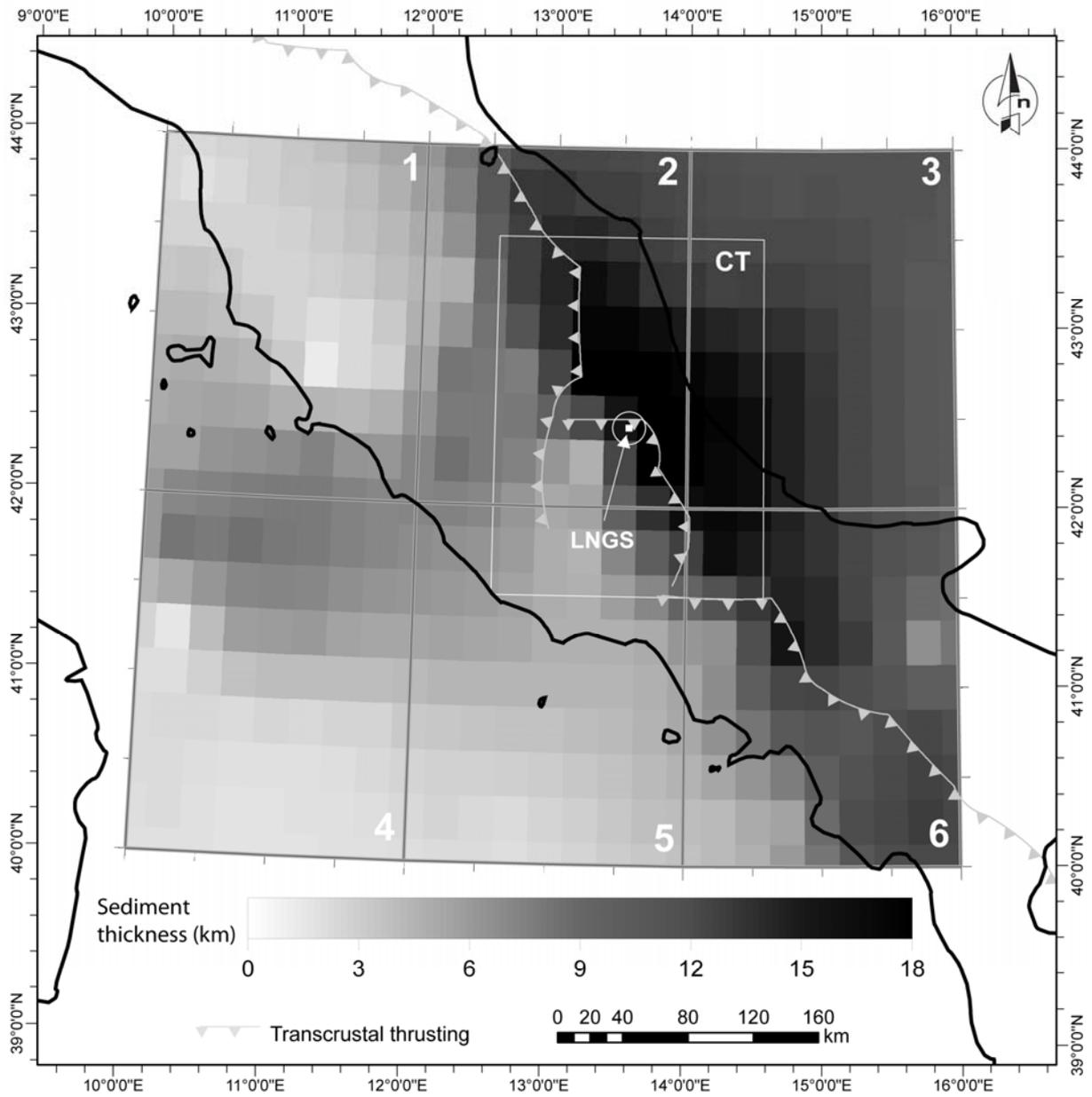

b) Upper Crust

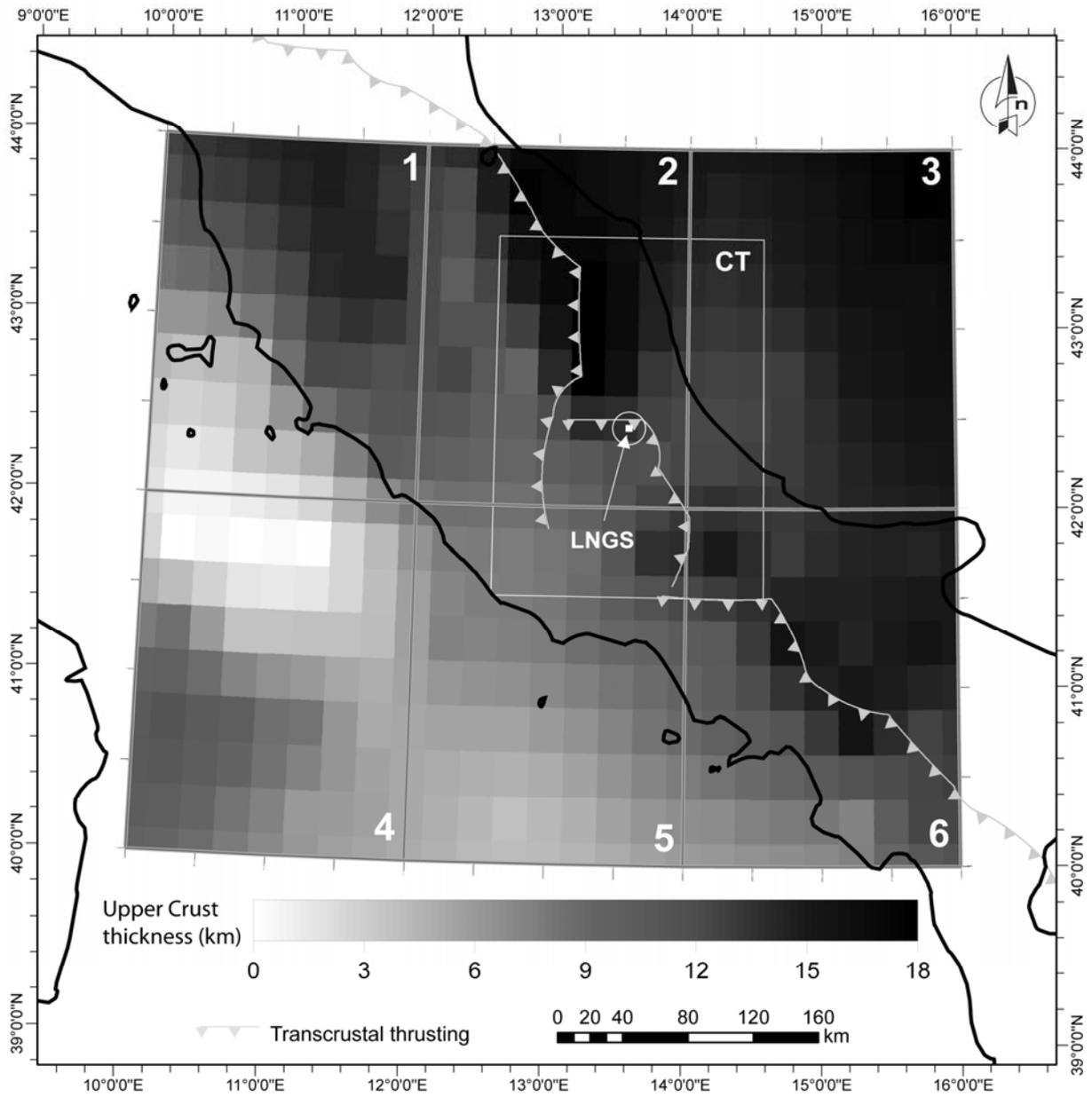

c) Lower Crust

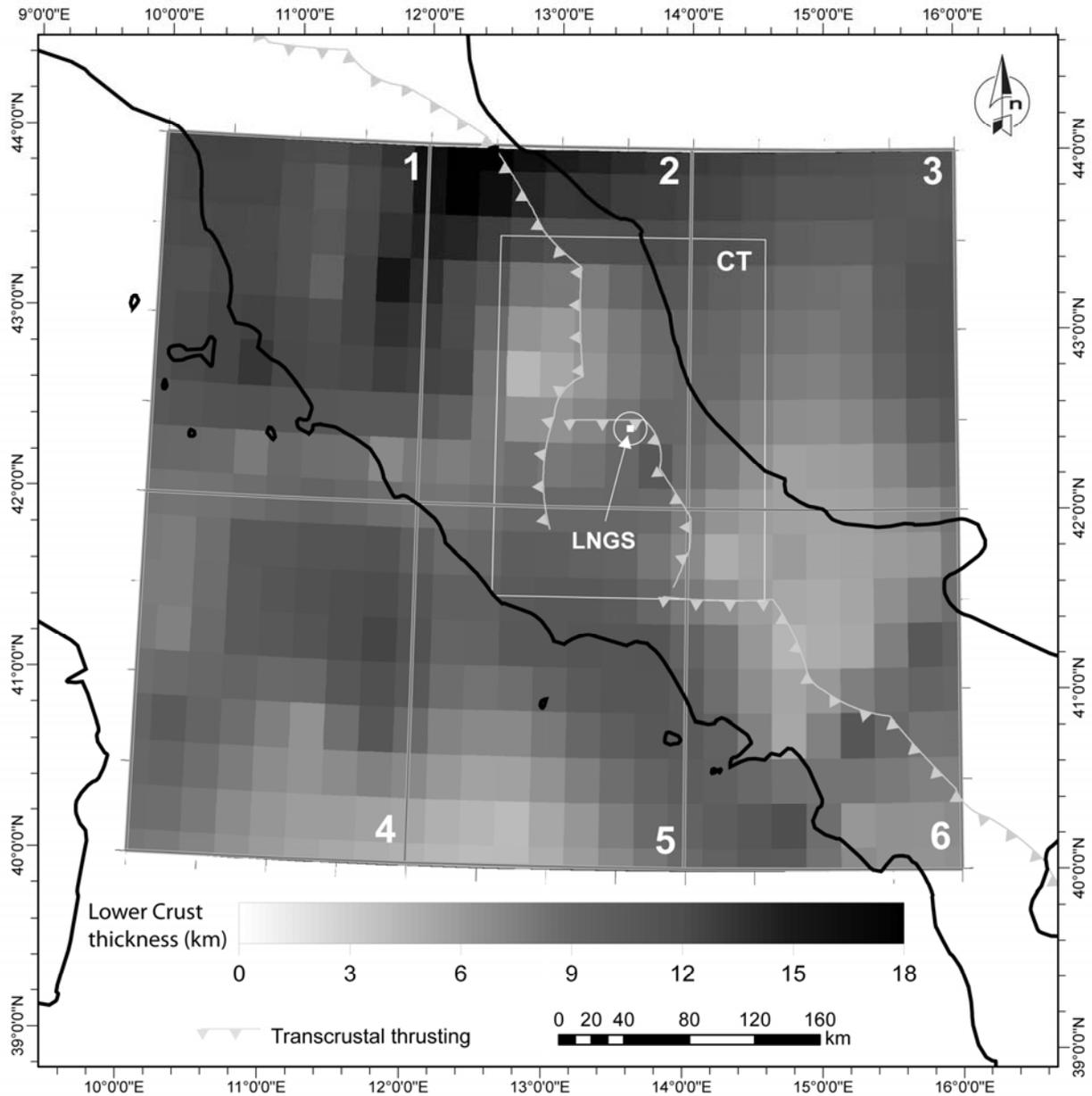

Fig. 5 – Thickness of the three layers according to the 3D reconstructed geological model. a) Sediments, b) Upper Crust, c) Lower Crust.